\documentclass[%
reprint,
 amsmath,amssymb,
aps,
floatfix,
]{revtex4-1}

\usepackage{graphicx}
\usepackage{dcolumn}
\usepackage{bm}
\usepackage{appendix}
\usepackage{subfig}
\usepackage{wrapfig}
\usepackage{verbatim}
\usepackage[export]{adjustbox}
\usepackage{float}
\def\dprod{\mathop{\displaystyle \prod }}
\def\func#1{\mathop{\rm #1}}
\def\limfunc#1{\mathop{\rm #1}}

\begin{document}


\title{Fluctuations and Correlations in a Model of Extreme Introverts and Extroverts}  

\author{Mohammadmehdi Ezzatabadipour\textsuperscript{1, 2}}
\author{Weibin Zhang\textsuperscript{1, 2}}
\author{Kevin E. Bassler\textsuperscript{1, 2}}%
\email{bassler@uh.edu }

\author{R. K. P. Zia\textsuperscript{3,4}}%

\affiliation{%
\textsuperscript{1}\textit{Physics Department, University of Houston, Houston, Texas 77204, USA}\\ \\
\textsuperscript{2}\textit{Texas Center for Superconductivity, University of Houston, Houston, Texas 77204, USA}\\ \\
\textsuperscript{3}\textit{Center for Soft Matter and Biological Physics, Department of Physics, Virginia Polytechnic Institute and State University, Blacksburg, Virginia 24061, USA}\\ \\
\textsuperscript{4}\textit{Department of Physics and Astronomy, UNC Asheville, North Carolina, 28804, USA}\\
}%



\begin{abstract}
Unlike typical phase transitions of first and second order, a system
displaying the Thouless effect exhibits characteristics of both at the
critical point (jumps in the order parameter and anomalously large
fluctuations). An $extreme$ Thousless effect was observed in a
recently introduced model of social networks consisting of `introverts and
extroverts' ($XIE$). We study the fluctuations and correlations of this
system using both Monte Carlo simulations and analytic methods based on a 
self-consistent mean field theory. Due to the symmetries in the model, we
derive identities between all independent two point correlations and
fluctuations in three quantities (degrees of individuals and the total
number of links between the two subgroups) in the stationary state. As
simulations confirm these identities, we study only the fluctuations in
detail. Though qualitatively similar to those in the 2D Ising model, there
are several unusual aspects, due to the extreme Thouless effect. All these
anomalous fluctuations can be quantitatively understood with our theory,
despite the mean-field aspects in the approximations. In our theory, we
frequently encounter the `finite Poisson distribution' (i.e., $x^{n}/n!$ for 
$n\in \left[ 0,N\right] $ and zero otherwise). Since its properties appear
to be quite obscure, we include an Appendix on the details and the related
`finite exponential series' $\sum_{0}^{N}x^{n}/n!$. Some simulation studies
of joint degree distributions, which provide a different perspective on
correlations, have also been carried out.
\end{abstract}

\maketitle

\section{\label{sec:Intro}Introduction}

For systems undergoing phase transition, the study of fluctuations and
correlations usually offer insight into the underlying collective behavior
of the constituents. Such studies are especially important for typical
second order transitions, where large, anomalous, and non-analytic
properties emerge. Somewhat outside the conventional wisdom of critical
phenomena are `mixed order transitions,' also known as the Thouless effect.
Displaying characteristics of both a first order transition (e.g., jumps in
the order parameter) and a second order one (e.g., anomalously large
fluctuations), this effect has been studied continuously \cite{kafri_why_2000, poland_phase_1966,fisher_effect_1966}
since fifty years ago, when Thouless' introduction of the inverse distance
squared Ising model \cite{Thouless_long-range_1969}. More recently, Bar and
Mukamel \cite{BarMukamel} coined the term `extreme Thouless effect' (ETE)
for a system that displays maximal jumps (e.g., $-1$ magnetisation to $+1$
across the transition) and `infinite' fluctuations (i.e., variations scaling
with the whole system) at the transition itself.

Though the ETE effect seems exotic, a natural route exists for constructing
an arguably trivial (and exactly solvable) system which displays it.
Introduced in the context of a kinetic Ising model \cite{FFK16}, it is best
viewed as a static system with a special Hamiltonian. Consider the
non-interacting Ising model with $\mathcal{N}$ spins ($s_{\alpha }=\pm 1$)
in an external field, $H$, at inverse temperature $\beta =1$. Thus $P\left(
M\right) $, the exact probability for finding the system with a given total
magnetisation $M$, is proportional to a binomial times $e^{HM}$, while the
Landau free energy, $-k_{B}T\ln P$, is $\ln \frac{\mathcal{N}+M}{2}!+\ln 
\frac{\mathcal{N-}M}{2}!-HM$ (apart from a const). All we needed to produce
a ETE is to introduce an interaction Hamiltonian which cancels the first two
terms here, i.e., $\mathcal{H}\left( \left\{ s_{i}\right\} \right) =-\ln %
\left[ \left( \mathcal{N}+\sum_{\alpha }s_{\alpha }\right) /2\right] !-\ln %
\left[ \left( \mathcal{N-}\sum_{\alpha }s_{\alpha }\right) /2\right] !$.
Now, the free energy (at $\beta =1$) is just $-HM$. Restricted to $m\equiv M/%
\mathcal{N}\in \left[ -1,1\right] $, the minimum is located at $sign\left(
H\right) $, so that $m$ literally jumps from $-1$ to $+1$ when $H$ crosses $%
0 $. Meanwhile, at $H=0$, $P$ is completely flat over the \textit{entire}
interval. Despite such drastic displays, there is no correlation between any
pair of spins. In this paper, we will explore a less trivial system which
displays the ETE, and in particular, the non-vanishing correlations between
its constituents.

The model system we study consists of a network of dynamic links, motivated
by the changing contacts between individuals in a social setting\cite%
{liu2013modeling}. In a typical population, we can expect to find a range of
preferences for how many contacts seems best, e.g., introverts preferring
few and extroverts, many. Labeling a node (an individual) by $i$ and its
preferences by $\kappa _{i}$, we evolve our simple model by choosing a node
at random and noting its degree $k_{i}$ -- the number of links (contacts) it
has. If $k_{i}<\kappa _{i}$, it chooses another node (which is not already
in contact) at random and makes a link to that. Otherwise, it chooses a
random existing link and cuts it. At any time, this social network is
completely described by the adjacency matrix $\mathbb{A}$, the elements of
which are binary: $a_{ij}=a_{ji}=1$ (or $0$) if the link between nodes $i$
and $j$ is present (or absent)\footnote{%
We exclude all self-contacts: $a_{ii}\equiv 0$.}. Thus, the evolution of our
system resembles that of a special kinetic two dimensional (2D) Ising model,
in which a pair of spins ($s_{ij}=2a_{ij}-1=s_{ji}$) can be flipped in each
step, depending on how many other spins in the same row (or column) are the
same. In general, the dynamics of such a system do not obey detailed balance%
\cite{liu2013modeling}, so that the system eventually settles into a
non-equilibrium steady state (NESS), characterized by a stationary
distribution, $\mathcal{P}^{\ast }\left( \mathbb{A}\right) $, as well as
persistent probability currents, $\mathcal{K}^{\ast }\left( \mathbb{%
A\rightarrow A}^{\prime }\right) $ \cite{zia2007probability}. Needless to
say, understanding the collective behavior of this system is extremely
challenging and so, only some simpler versions have been studied in detail
so far. Notable examples are (a) a homogeneous population, with just one $%
\kappa $ for all \cite{liu2013modeling}, (b) a population with $N_{I}$
`introverts' and $N_{E}$ `extroverts,' specified by $\kappa _{I}<\kappa _{E}$
\cite{liu2014modeling}, and most extensively, (c) the $XIE$ model \cite%
{zia2012extraordinary, liu2013extraordinary, bassler2015extreme,
bassler2015networks, ZZEB2018} for the extreme case of the latter, with $%
\kappa _{I}=0$ and $\kappa _{E}=\infty $. In particular, the ETE was found
in the last case \cite{zia2012extraordinary, liu2013extraordinary}. Beyond
these models with static $\kappa $'s, the further motivation is to exploit
adaptive networks\footnote{%
There is considerable work on adaptive networks in the literature,
especially in connection with epidemic spreading\cite{GrossGeneral}.
However, nearly all other approaches are based on rewiring existing links
from, e.g., an infected node to a healthy one. By introducing \textit{%
preferred degrees}, we believe our models capture human behavior more
realistically, as well as the possibility of diverse and inhomogeneous
population.}, in which the preference may be dynamic due to personal reasons
or external controls, to model realistic social phenomena such as revelation
of hidden secrets or response to epidemics\cite{platini2010, jolad2012}.

This paper reports the continuing study of the $XIE$ model. Specifically,
despite the apparent presence of long-ranged and multi-spin interactions\cite%
{zia2012extraordinary, liu2013extraordinary}, mean field approximations
(MFA) have been extraordinarily successful in capturing all aspects of the
ETE\cite{bassler2015extreme, bassler2015networks, ZZEB2018}. A series of
natural questions thus arises: What is the nature of the correlations? Are
they small, so that the MFAs are so effective? Given that intimate relations
exist between correlations and fluctuations, how can they be small when the
fluctuations are `extreme' at the critical point? Before attempting to
answer these questions, we provide details of the model in the next Section,
as well as the similarities and contrasts with the standard 2D Ising model.
Turning to fluctuations and correlations in the following Section, we
display the identities which relate fluctuations of magnetisation-like
quantities to the standard two-point correlations, and we introduce another
measure of correlations which is natural for networks but not for Ising
systems. Stimulation data and various mean-field approaches will be
presented next. Much of the theory is based on \textit{finite} Poisson
distributions, with somewhat unusual properties. Since they appear to be
rarely discussed in the literature, we include a substantial section in the
Appendix on the results we derived. We end with a summary and outlook.

\section{\label{sec:Models}The $XIE$ model and connections to the Ising model}

\begin{figure}[ht]
\includegraphics[width=0.35\textwidth]{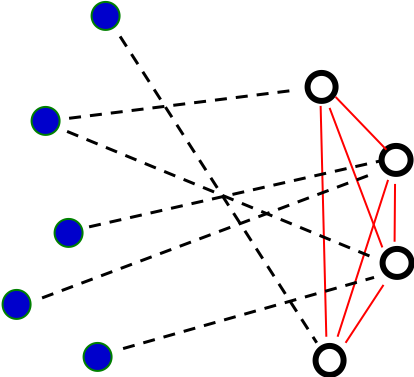}
\vspace{0.2cm}
\caption{A typical steady state configuration of an $N_{I,E}=5,4$
system. The $I$'s are represented as solid (blue on line) circles while $E$%
's are open ones. The $i$,$\eta $ labels nodes from top to bottom. All $I$-$%
I $ ($E$-$E$) links are absent (present, solid lines, red on line), while
only the $I$-$E$ links (dashed lines) are dynamic.}
\label{pic}
\end{figure}

The dynamic network we study here, the $XIE$ model, is an extreme version of
social connections between $N_{I}$ introverts ($I$'s) and $N_{E}$ extroverts
($E$'s). The system evolves in discrete time steps: At each step, one of the 
$N_{I}+N_{E}$ nodes is randomly chosen to act. If an $I$ with $k$ ($>0$)
links is chosen, it will cut one of the links with probability $1/k$. If an $%
E$ with no connections to $p$ others is chosen, it will make a connection to
one of these with probability $1/p$. Since no $I$ makes links and no $E$
cuts them, it is clear that, starting with any initial network, this minimal
system will quickly settle into two distinct subgroups: $I$'s with no links
amonst themselves and a complete graph of all the $E$'s. The only dynamic
links are the cross links between these groups and in this sense, the
fluctuating part of our network ranges over all the \textit{bipartite}
graphs. Thus, we only need to focus on one of the off diagonal blocks of the
adjacency matrix $\mathbb{A}$, i.e., the incidence matrix $\mathbb{N}$. Let
us denote the elements of $\mathbb{N}$ by $n_{i\eta }$, with $i=1,...,N_{I}$
and $\eta =1,...,N_{E}$, which assumes the values $1$ and $0$ if the link
between introvert $i$ and extrovert $\eta $ is present or absent,
respectively\footnote{%
Unless stated otherwise, Latin/Greek indices are reserved for
introverts/extroverts.}. In the sense that $n$ is an occupation variable, we
will also use the language of \textit{particle/hole} for these states. A
typical configuration of an $N_{I,E}=5,4$ system in its steady state is
illustrated in Fig.\ref{pic}, for which we have%
\begin{equation*}
\mathbb{N}=\left( 
\begin{array}{cccc}
0 & 0 & 0 & 1 \\ 
1 & 0 & 1 & 0 \\ 
0 & 1 & 0 & 0 \\ 
0 & 1 & 0 & 0 \\ 
0 & 0 & 1 & 0%
\end{array}%
\right)
\end{equation*}%

Although our focus
here will be much more modest, the ultimate goal of the statistical
mechanics of $XIE$ model is the solution to the master equation for the
probability distribution 
\begin{equation*}
\mathcal{P}\left( \mathbb{N};t+1\right) -\mathcal{P}\left( \mathbb{N}%
;t\right) =\sum_{\mathbb{N}^{\prime }}\mathbb{\mathcal{L}}\left( \mathbb{%
N\leftarrow N}^{\prime }\right) \mathcal{P}\left( \mathbb{N}^{\prime
};t\right)
\end{equation*}%
which governs the evolution of $\mathbb{N}$ in discrete time, given an
initial distribution $\mathcal{P}\left( \mathbb{N};0\right) $. Here, $%
\mathcal{L}$ is known as the Liouvillian and plays the role of the
Hamiltonian in the Schr\"{o}dinger equation. It is composed of the
transition probabilities (from $\mathbb{N}^{\prime }$ to $\mathbb{N}$) and,
since its explicit form is quite involved but not relevant, will not be
discussed further. Nevertheless, we emphasize that it is known to obey
detailed balance \cite{zia2012extraordinary, liu2013extraordinary} and so,
the system eventually settles into an equilibrium stationary state with
distribution\footnote{%
We denote quantities associated with the stationary state by a superscript ($%
^{\ast }$).} $\mathcal{P}^{\ast }\left( \mathbb{N}\right) \propto \Pi
_{i}\left( k_{i}!\right) \Pi _{\eta }\left( p_{\eta }!\right) $ (and zero
persistent probability currents \cite{zia2007probability}). Here,%

\begin{equation*}
k_{i}=\sum_{\eta }n_{i\eta };~~p_{\eta }=N_{I}-\sum_{i}n_{i\eta }
\end{equation*}%

denote, respectively, the particle and hole content of a row and column.
Note that $k_{i}$ and $\left( N_{I}-p_{\eta }\right) $ are just the degrees
of nodes $i$ and $\eta $. Since an $I$ ($E$) prefers to have no links (links
to all), $k$ ($p$) is a measure of the `frustration' of the node. Finally,
given that $\mathcal{P}^{\ast }$ is like a Boltzmann factor, we may regard%
\begin{equation}
\mathcal{H}\left( \mathbb{N}\right) =-\sum_{i}\ln \left( k_{i}!\right)
-\sum_{\eta }\ln \left( p_{\eta }!\right)  \label{Ham}
\end{equation}%
as a Hamiltonian (with inverse temperature $\beta =1$ in this case).

As with $\mathbb{A}$ above, $\mathbb{N}$ can be regarded as a 2-D Ising
model in the lattice gas representation. Unlike $\mathbb{A}$, which must
remain symmetric to model undirected links in a network, there are no
constraints on $\mathbb{N}$, so that the dynamics involve only single `spin'
flips. Of course, there are major differences between $XIE$ model and the
Ising, some of which are listed here. The only (explicit) control parameters
in our minimal model are $N_{I,E}$. While $\mathcal{N\equiv }N_{I}N_{E}$
corresponds to the overall system size of the Ising model, the aspect ratio, 
$N_{I}/N_{E}$, is rarely considered as a variable. Alternatively, we often
use the average and difference%
\begin{equation}
N\equiv \left( N_{I}+N_{E}\right) /2;~~\Delta \equiv N_{E}-N_{I}
\label{N+Delta}
\end{equation}%
as parameters. There is no spatial structure in our network; instead the
system is symmetric under permutation of any row and any column (interchange
between pairs of $I$-'s or $E$'s). Thus, boundary conditions are irrelevant
here. Meanwhile, the $XIE$ dynamics corresponds to simple spin flip in
Ising, as it stipulates (i) choosing a row or a column at random, (ii)
flipping a random $1$ to $0$ in the chosen row, and (iii) flipping a random $%
0$ to $1$ in the chosen column. From these rules, we see that, if $\Delta >0$
say, then more attempts will be made to flip from $0$ to $1$, so that $%
\Delta $ can be regarded as an external magnetic field in the Ising model.
Thus, the Ising symmetry corresponds to interchanging $1\Leftrightarrow 0$ 
\textit{together with }$\Delta \Leftrightarrow -\Delta $. It is possible to
introduce a further bias favoring say, the choice of an $E$ to act. Such a
bias would mimic an addition field-like parameter, $H$. Similarly, we may
introduce a temperature-like variable, $\beta $, and perform simulations
with the Boltzmann factor $\exp \left\{ -\beta \mathcal{H}\right\} $. Though
our main focus is $\beta -1=H=0$ here, we will mention briefly in the last
section explorations that parallel those of the Ising model: in the full $%
\beta $-$H$ plane while keeping $\Delta =0$. In this context, we are
studying a 2-D Ising model with `long ranged,' multi-spin interactions for a
Hamiltonian. Indeed, every spin interacts with all other spin in the same
row or column\footnote{%
For example, $\ln \left( n_{1}+n_{2}+n_{3}\right) !=\left(
n_{1}n_{2}+n_{2}n_{3}+n_{3}n_{1}\right) \ln 2+n_{1}n_{2}n_{3}\ln \left(
3/4\right) $.}. Thus, the first impression is that correlations must be
quite serious, in the sense that it cannot be exponential (or algebraic)
decaying, as typical in the 2D Ising case. On the other hand, they cannot be
arbitrarily strong, since correlations are related to fluctuations. In the
next section, we will explore these relations in detail. Here, let us
provide a brief summary of the remarkable phenomena associated with the ETE.

\begin{figure}[ht]
\centering
\includegraphics[width=0.5\textwidth]{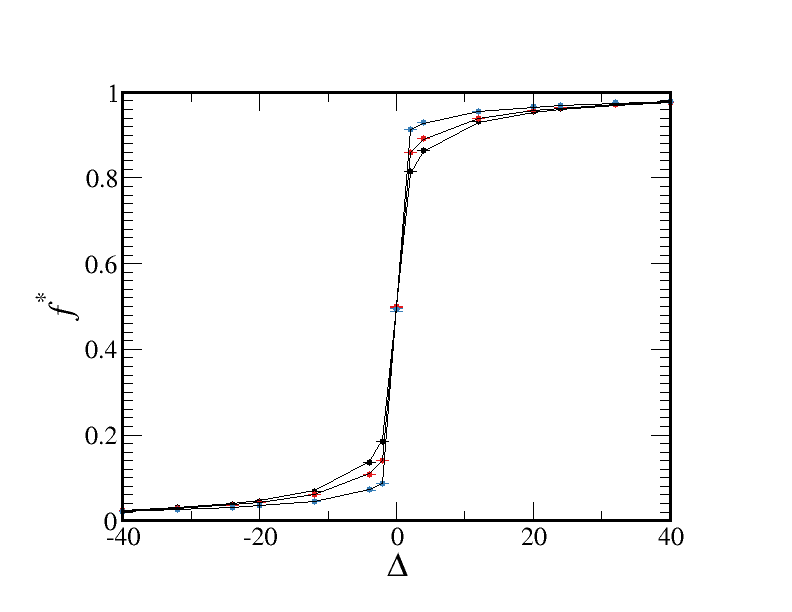}
\caption{The fraction of
cross-links in the steady state, $f^{\ast }$, as a function of $\Delta $.
The simulation data (symbols) for $N=40,100,400$ are shown along with the
theoretical predictions (denoted as $\tilde{f}$ in Section IV B
(lines). In the large $N$ limit, $f^{\ast }\left( \Delta \right) $
approaches the Heaviside step function, $\Theta \left( \Delta \right) $.}
\label{f-vs-Delta}
\end{figure}

Of the many macroscopic quantities of interest in our system, the simplest
is the total number of cross-links, or equivalently, the fraction of such
links:%
\begin{equation*}
X\equiv \sum_{i,\eta }n_{i\eta };~~f\equiv X/\mathcal{N}
\end{equation*}%
These correspond to the total net spin $M=\sum_{i,j}s_{ij}$ and the
magnetisation $m\equiv M/\mathcal{N}$ in the Ising model. The stationary
average\footnote{%
To be consistent with our notation, we should write $\left < ... \right > ^{\ast }$ for stationary averages. But, for the sake of
simplicity, we drop the superscript, since all averages considered in this
paper will be in the stationary state. By contrast, note that, e.g., while $%
f^{\ast }\equiv \left< f\right> $ is a single number, $f$
denotes a variable.}, $\left\langle X\right\rangle $, provides a coarse view
of how connected the network is, while the variance $\sigma
_{X}^{2}=\left\langle X^{2}\right\rangle -\left\langle X\right\rangle ^{2}$
is a measure of its fluctuations. Their analogs in the Ising model would be
the order parameter and susceptibility. The extraordinary behavior of these
quantities was first discovered \cite{zia2012extraordinary,
liu2013extraordinary} through simulations of systems with $N=100$ and a few $%
\Delta $'s in $\left[ -50,+50\right] $. Specifically, $f^{\ast }\equiv
\left\langle f\right\rangle $ jumps from about $14\%$ to $86\%$ when $\Delta 
$ is tuned to $-2$ (i.e., $101$ introverts and $99$ extroverts) or $+2$. In
the Ising language, the jump in $\left\langle m\right\rangle $ would be from 
$-0.7$ to $+0.7$! This behavior is illustrated in Fig. \ref{f-vs-Delta}, for 
$N=40$ and $400$ as well. 

Meanwhile, when $f$ is monitored in the `critical' system ($\Delta
_{c}=0$), it is found to execute an \textit{unbiased random walk} \cite%
{bassler2015extreme} between `soft walls' at $f_{0}\sim 0.2$ and $1-f_{0}$.
In other words, its stationary distribution, $P^{\ast }\left( f\right) $,
resembles a mesa which spans nearly the entire allowed interval $\left[ 0,1%
\right] $. In subsequent simulation and theoretical studies \cite%
{bassler2015extreme, bassler2015networks, ZZEB2018}, we found that, as $%
N\rightarrow \infty $, the jump becomes maximal, while $f_{0}$ vanishes
asymptotically as $\sqrt{\left( \ln N^{2}\right) /N}$. To emphasize, the
latter means $P^{\ast }\left( f\right) \rightarrow 1$ for all $f\in \left[
0,1\right] $, while $\sigma _{X}^{2}\rightarrow \mathcal{N}^{2}/12$! While
such extreme features are the defining signatures of an ETE, this work is
devoted to exploring the implications of such gargantuan fluctuations for
the correlations between the links in our network.

\section{\label{sec:CorrFluc}Correlations and fluctuations}

The simplest measure of correlations in the 2D Ising model is the two-point
function, $\left\langle s_{ij}s_{k\ell }\right\rangle -\left\langle
s_{ij}\right\rangle \left\langle s_{k\ell }\right\rangle $,\ the integral
(sum) of which is the variance $\left\langle M^{2}\right\rangle
-\left\langle M\right\rangle ^{2}$, a measure of the fluctuations. The
analogs in $XIE$ are straightforward and, thanks to the permutation
symmetry, they are relatively easy to compute (within the MFAs). In the
analysis, we will be led naturally to the degree distributions of the nodes.
Though, standard in the study of networks, their analogs in Ising model have
rarely been examined. Beyond these, we will consider another natural measure
of correlation for networks, the \textit{joint distribution} of degrees of
two nodes. Though easily measured in simulations, understanding their
behavior remains a challenge.

\subsection{\label{sec:2pf}Two point functions}

We begin with the fluctuation-correlation identities, the foremost of which
is simplest is 
\begin{equation*}
\sigma _{X}^{2}=\sum_{i,\eta ,j,\gamma }\left\langle n_{i\eta }n_{j\gamma
}\right\rangle -\left\langle X\right\rangle ^{2}
\end{equation*}%
Unlike the Ising case, ours is much simpler, since permutation symmetry of
the steady state implies that there are only three distinct correlations. As 
$n_{i\eta }n_{i\eta }=n_{i\eta }$, we need to focus on cases where only one (%
$I$ or $E$) set of indices \textit{differ}, or both $I$ and $E$ indices
differ. Thus, we define the correlations, for \textit{any} $i\neq j$ and $%
\eta \neq \gamma $,%
\begin{eqnarray*}
\chi _{I} &\equiv &\left\langle n_{i\eta }n_{j\eta }\right\rangle -\left(
f^{\ast }\right) ^{2} \\
\chi _{E} &\equiv &\left\langle n_{i\eta }n_{i\gamma }\right\rangle -\left(
f^{\ast }\right) ^{2} \\
\chi _{IE} &\equiv &\left\langle n_{i\eta }n_{j\gamma }\right\rangle -\left(
f^{\ast }\right) ^{2}
\end{eqnarray*}%
since $\left\langle n_{i\eta }\right\rangle =\left\langle X\right\rangle /%
\mathcal{N}=f^{\ast }$. As a result, instead of the Ising relation, we find
a much simpler fluctuation-correlation identity:%
\begin{equation}
\frac{\sigma _{X}^{2}}{\mathcal{N}}=\chi _{0}+\left( N_{I}-1\right) \chi
_{I}+\left( N_{E}-1\right) \chi _{E}+\left( N_{I}-1\right) \left(
N_{E}-1\right) \chi _{IE}  \label{FCid}
\end{equation}%
where%
\begin{equation}
\chi _{0}\equiv f^{\ast }\left( 1-f^{\ast }\right)  \label{chi0}
\end{equation}%
is the variance of any single link: $\left\langle n_{i\eta
}{}^{2}\right\rangle -\left\langle n_{i\eta }\right\rangle ^{2}$. Of course,
we can consider normalized correlations $\chi _{I,E,IE}/\chi _{0}$. But, as
will be shown below, these present unnecessary theoretical challenges when
we study the critical system.

There are three unknown $\chi $'s on the right of Eqn. (\ref{FCid}). All can
be determined in terms of fluctuations if we consider the degree
distributions. In the stationary state, only two such distributions can be
distinct: one associated with any $I$ and the other, with any $E$. To be
precise, we write\footnote{%
Again, we drop the superscript $^{\ast }$ for $\rho $ for simplicity,
despite that it is a steady state quantity.}%
\begin{equation*}
\rho _{I}\left( k\right) \equiv \sum_{\mathbb{N}}\delta \left( k-\sum_{\eta
}n_{i\eta }\right) \mathcal{P}^{\ast }\left( \mathbb{N}\right)
\end{equation*}%
where $\delta $ is the Kronecker delta. Of course, we can write a similar
expression for $\rho _{E}$. But, for symmetry reasons, it is better to
consider the `\textit{hole-distribution}'%
\begin{equation*}
\zeta _{E}\left( p\right) \equiv \sum_{\mathbb{N}}\delta \left(
p-N_{I}+\sum_{i}n_{i\eta }\right) \mathcal{P}^{\ast }\left( \mathbb{N}\right)
\end{equation*}%
From these, we study the averages and variances\footnote{%
Note the bar above denote averages over the degree distributions, not the
microscopic $\mathcal{P}\left( \mathbb{N}\right) $.}%
\begin{eqnarray*}
\bar{k} &\equiv &\sum_{k}k\rho _{I}\left( k\right) ;~~\sigma _{k}^{2}\equiv 
\overline{k^{2}}-\bar{k}^{2} \\
\bar{p} &\equiv &\sum_{p}p\zeta _{E}\left( p\right) ;~~\sigma _{p}^{2}\equiv 
\overline{p^{2}}-\bar{p}^{2}
\end{eqnarray*}%
Focusing on the $I$'s for now, we find an expected result%
\begin{equation*}
\bar{k}=\sum_{\mathbb{N}}\left( \sum_{\eta }n_{i\eta }\right) \mathcal{P}%
^{\ast }\left( \mathbb{N}\right) =N_{E}f^{\ast }
\end{equation*}%
which is also $\left\langle X\right\rangle /N_{I}$. Meanwhile,%
\begin{eqnarray*}
\overline{k^{2}} &=&\sum_{k}k^{2}\rho _{I}\left( k\right) =\left\langle
\sum_{\eta }n_{i\eta }\sum_{\gamma }n_{i\gamma }\right\rangle \\
&=&N_{E}\left[ \left\langle n_{i\eta }\right\rangle +\left( N_{E}-1\right)
\left\langle n_{i\eta }n_{i\gamma }\right\rangle \right]
\end{eqnarray*}%
so that $\sigma _{k}^{2}=N_{E}\left[ f^{\ast }+\left( N_{E}-1\right)
\left\langle n_{i\eta }n_{i\gamma }\right\rangle -N_{E}\left( f^{\ast
}\right) ^{2}\right] $, and we arrive at a `fluctuation-correlation identity
for introverts': 
\begin{equation}
\frac{\sigma _{k}^{2}}{N_{E}}=\chi _{0}+\left( N_{E}-1\right) \chi _{E}
\label{iFCid}
\end{equation}%
To be precise, we should state that there is an exact relationship between
the variance of an $I$'s degree and the correlation of two of its links (to
two different $E$'s). Since $\sigma _{p}^{2}$ is also the variance of the
extroverts' degree distribution, we easily find a similar identity for the
extroverts:%
\begin{equation}
\frac{\sigma _{p}^{2}}{N_{I}}=\chi _{0}+\left( N_{I}-1\right) \chi _{I}
\label{eFCid}
\end{equation}%
Thus, \textit{all} the $\chi $'s are known, once we obtain the variances: $%
\sigma _{k}^{2}$, $\sigma _{p}^{2}$, and $\sigma _{X}^{2}$. Note that,
though we may consider similar quantities in the 2D Ising model, there is
typically little reason to study the row- or column-magnetisation, i.e., the
sums of the spins in a row or a column. Nevertheless, they do play crucial
roles in highly anisotropic systems, such as driven diffusive systems\cite%
{SZ95, DZ18} where the order parameter is the row (or column) magnetisation.
A further note concerns $\sigma ^{2}/N$ in Eqns.(\ref{FCid},\ref{iFCid}):
For non-critical systems, we expect `normal' fluctuations and these to be $%
O\left( 1\right) $ in the thermodynamic limit, so that $\left(
N_{I,E}-1\right) \chi _{I,E}$ and $\left( N_{I}-1\right) \left(
N_{E}-1\right) \chi _{IE}$ should be good scaling variables.

There is another perspective of these identities which provides us with a
gauge on how inter-dependent our variables are. In particular, if the degree
of the $I$'s (and $E$'s) were iid's from the same $\rho _{I}$ (and $\zeta
_{E}$), then we would find%
\begin{equation}
\sigma _{X}^{2}=N_{I}\sigma _{k}^{2}=N_{E}\sigma _{p}^{2}  \label{iidVar}
\end{equation}%
However, Eqns. (\ref{iFCid},\ref{eFCid}) can be exploited to recast Eqns (%
\ref{FCid}) in several equivalent forms%
\begin{eqnarray}
\sigma _{X}^{2} &=&N_{I}\sigma _{k}^{2}+\mathcal{N}\left( N_{I}-1\right) %
\left[ \chi _{I}+\left( N_{E}-1\right) \chi _{IE}\right]  \label{test-iid1}
\\
&=&N_{E}\sigma _{p}^{2}+\mathcal{N}\left( N_{E}-1\right) \left[ \chi
_{E}+\left( N_{I}-1\right) \chi _{IE}\right]  \label{test-iid2}
\end{eqnarray}%
Applied to Erd\H{o}s-R\'{e}nyi graphs \cite{E-R} with average $f^{\ast }$,
all these $\chi $'s vanish and Eqn. (\ref{iidVar}) is verified. In $XIE$,
simulations show that these $\chi $'s are non-trivial and differences like $%
\sigma _{X}^{2}-N_{I}\sigma _{k}^{2}$ are significant.

We end this subsection with the analysis of the full stationary distribution
of $X$ \footnote{%
To be clear, the full notation should be denoted $P^{\ast }\left( X;N,\Delta
\right) $, showing the dependence on the control parameters $\left( N,\Delta
\right) $. For simplicity, we suppress the latter except when their presence
are crucial.}

\begin{equation}
P^{\ast }\left( X\right) \equiv \sum_{\mathbb{N}}\delta \left(
X-\sum_{i,\eta }n_{i\eta }\right) \mathcal{P}^{\ast }\left( \mathbb{N}\right)
\label{P*(X)}
\end{equation}%
and the related \textit{fixed} $X$ \textit{ensembles}, which are just
cross-sections of $\mathcal{P}^{\ast }\left( \mathbb{N}\right) $ with a
given $X$. They will play key roles in the analysis of the critical system.
Their analogs in the Ising model, fixed $M$ ensembles, have received much
attention for both physical and mathematical reasons. Many systems in nature
with conserved $M$, e.g., liquid-vapor and binary alloys, are formulated in
this manner and typically presented as the `Ising lattice gas' \cite%
{YangLee52}. Such systems allow us to explore a variety of phenomena, e.g.,
phase co-existence, interfacial properties, nucleation, metastability and
hysteresis. Mathematically, since fixed $M$ ensembles are conjugate to
fluctuating $M$ ensembles with an external $H$, the associated free energies
are Legendre transforms of each other (in the thermodynamic limit). Thus,
they offer different approaches to analyze subtle singularities in the free
energy, such as those associated with metastability. For the $XIE$ model, we
discovered that fixed $X$ ensembles offer sufficient insight for
constructing a mean-field approach that provides spectacular agreement with 
\textit{all} simulation data, including those for the critical system\cite%
{ZZEB2018}. To avoid confusion, we will denote averages in such ensembles
with extra caret above, e.g.,%
\begin{equation}
\widehat{\left\langle \mathcal{O}\left( \mathbb{N}\right) \right\rangle }%
\equiv \sum_{\mathbb{N}}\mathcal{O}\left( \mathbb{N}\right) \delta \left(
X-\sum_{i,\eta }n_{i\eta }\right) \mathcal{P}^{\ast }\left( \mathbb{N}\right)
\end{equation}%
with the understanding that $X$ (alternatively, $f\equiv X/\mathcal{N}$) is
a control parameter here, with no fluctuations. Thus, for example, 
\begin{equation}
\hat{\chi}_{0}=f\left( 1-f\right)
\end{equation}%
is just a given constant, unlike $\chi _{0}$ in (\ref{chi0}) which must be
computed from $\left( N,\Delta \right) $. Clearly, for such ensembles, $%
\widehat{\sigma _{X}^{2}}\equiv 0$ and Eqn. (\ref{FCid}) reduces to%
\begin{equation}
0=\hat{\chi}_{0}+\left( N_{I}-1\right) \hat{\chi}_{I}+\left( N_{E}-1\right) 
\hat{\chi}_{E}+\left( N_{I}-1\right) \left( N_{E}-1\right) \hat{\chi}_{IE}
\label{FCid-fX}
\end{equation}%
Thus, for fixed $X$ ensembles, some correlations must be negative. As will
be shown below, these $\hat{\chi}$'s are crucial for understanding the
extraordinary correlations in the critical system ($\Delta =0$).

In the next Section, we will present simulation data and MFA's for
understanding them.

\subsection{\label{sec:JDD1} Correlation in joint degree distributions}

Beyond the two point function, there is a seemingly infinite variety of
other possible measure of correlations. Here, we focus on one other quantity
which appears natural for networks, namely, the joint degree distribution.
In general, if $x$ and $y$ are independent variables distributed according
to $\rho _{x}\left( x\right) $ and $\rho _{y}\left( y\right) $, then the
joint distribution $\rho \left( x,y\right) $ is just the product $\rho
_{x}\left( x\right) \rho _{y}\left( y\right) $. Thus, the difference between
them is a good measure of the correlation between $x$ and $y$. In our case,
we can study three such distributions: two $I$'s ($\rho \left( k,k^{\prime
}\right) $), two $E$'s ($\zeta \left( p,p^{\prime }\right) $), and one of
each ($\mu \left( k,p\right) $). To save notation, we will use, as above, $k$
for the degree associated with an $I$ and $p$ for the `\textit{hole-degree}'
associated with an $E$. Thus, e.g., 
\begin{equation}
\mu \left( k,p\right) =\sum_{\mathbb{N}}\delta \left( k-\sum_{\eta }n_{1\eta
}\right) \delta \left( p-N_{I}+\sum_{i}n_{i1}\right) \mathcal{P}^{\ast
}\left( \mathbb{N}\right)  \label{mu}
\end{equation}%
is the joint distribution for an $I$ and an $E$. In the steady state, all
nodes in each group should be equivalent and so, we write $1$'s in the above
for convenience. The differences, such as%
\begin{equation}
\mu \left( k,p\right) -\rho \left( k\right) \zeta \left( p\right)
\end{equation}%
are measures of the correlations between the two nodes. Finding a viable
theory to provide quantitative insights into these quantities has been
elusive. Below, we will only show simulation data and make some qualitative
statements. In this context, we will present the `normalized' differences,
e.g., 
\begin{equation}
C_{IE}\left( k,p\right) \equiv \frac{\mu \left( k,p\right) }{\rho \left(
k\right) \zeta \left( p\right) }-1  \label{C-def}
\end{equation}

Before proceeding to the data, we should emphasize that, though $\rho \left(
k,k^{\prime }\right) $ and $\rho \left( p,p^{\prime }\right) $ reduce to the
respective products for random $\mathbb{N}$'s (i.e., Erd\H{o}s-R\'{e}nyi
bipartite graphs), this is not the case for the mixed distribution $\mu $.
The reason is that, for any $I$-$E$ pair, their degrees are not entirely
independent, as both $k$ and $p$ are affected (oppositely) by the one link
between them. In particular, $n_{11}$ appears in both $\delta $ functions in
Eqn. (\ref{mu}), so that the sum does not factorize into the product of sums
over $\mathcal{P}^{\ast }$, even if $\mathcal{P}^{\ast }$ itself factorizes.
Deferring details of this analysis to an Appendix, we only quote the result
here\footnote{%
The superscript, $ER$, is remind us that this result only holds for Erd\H{o}%
s-R\'{e}nyi graphs.}%
\begin{equation}
C_{IE}^{ER}\left( k,p\right) =\left( \frac{k}{\bar{k}}-1\right) \left( 1-%
\frac{p}{\bar{p}}\right)  \label{C-ER}
\end{equation}%
where $\bar{k}/N_{E}=1-\bar{p}/N_{I}$ is the probability of any element in $%
\mathbb{N}$ being unity and illustrated in Fig \ref{JDD-ER}. Being a simple
quadrupole\footnote{%
If we write $\left( k-\bar{k}\right) =\bar{k}\cos \theta $ and $\left( p-%
\bar{p}\right) =\bar{p}\sin \theta $, then we find the standard quadrupole
form: $C\propto \sin 2\theta $.} in the $k$-$p$ plane, this scaling form
clearly displays the anti-correlation between the particle- and the
hole-count (i.e., one more $1$ in a row being correlated with one less $0$
in a column).

\section{\label{sec:Data+MFT}Simulation results and mean field approaches}

This Section is devoted to simulation studies and our understanding through
mean field approximations. For the two point correlations and fluctuations
(variances in the degrees of a node or the total number of cross-links), we
considered systems with $N=40,100$, and $400$, performing 10 independent
runs of $10^{7} \times \mathcal{N}$ MCS for each data point, where $\mathcal{N}=N_I \times N_E$. To obtain the averages
in the steady state, $\left\langle ...\right\rangle $, we discard the first
$10^{5} \times \mathcal{N}$ MCS and take measurement every $\mathcal{N}$ MCS. Following
earlier studies of our model \cite{liu2013extraordinary,bassler2015extreme},
we first use $\Delta $ as a control parameter, in the range $\left[ -40,40%
\right] $. Thanks to $I$-$E$ symmetry, the degrees of an $I$, $\rho
_{I}\left( k\right) $, in the stationary state is the same as the
distribution of holes, $\zeta _{E}\left( p\right) $, for an $E$ at the
opposite $\Delta $. Though not display here, we have explicitly verified
that such symmetries hold. Thus, in the following, we will consider on only $%
\rho \left( k\right) $, where we also dropped the subscript for simplicity.
This symmetry also implies that the stationary distribution of cross-links, $%
P^{\ast }\left( X\right) $, obeys%
\begin{equation*}
P^{\ast }\left( X;N,\Delta \right) =P^{\ast }\left( \mathcal{N}-X;N,-\Delta
\right)
\end{equation*}%
so that we can focus only on results for, say, $\Delta \leq 0$. Now, as Fig. %
\ref{f-vs-Delta} shows, as we study larger and larger systems with $\Delta
\neq 0$, we have access to a more and more limited region of $f$. On the
other hand, from Ref. \cite{ZZEB2018}, we see that a critical system behaves
like a superposition of fixed $f$ ones -- from $f_{0}$ to $1-f_{0}$. Thus,
we can access the intermediate $f$ regime by using fixed $f$ (or $X$)
ensembles. To generate such ensembles by Monte Carlo simulations, we begin
with a random $\mathbb{N}$ consisting of precisely $X$ `particles' and carry
out the standard Kawasaki particle-hole pair exchanges according to
Metropolis rates (i.e., accepting the attempt with probability $\min \left\{
1,e^{-\mathcal{H}}\right\} $ where $\delta \mathcal{H}$ is the difference in 
$\mathcal{H}$ as a result of the exchange). To obtain $\widehat{\left\langle
...\right\rangle }$, we discard the first $10^5 \times \mathcal{N}$ MCS and take
measurement every $\mathcal{N}$ MCS over $10$ independent runs.

\begin{figure*}
\begin{tabular}{cc}
 \includegraphics[width=80mm]{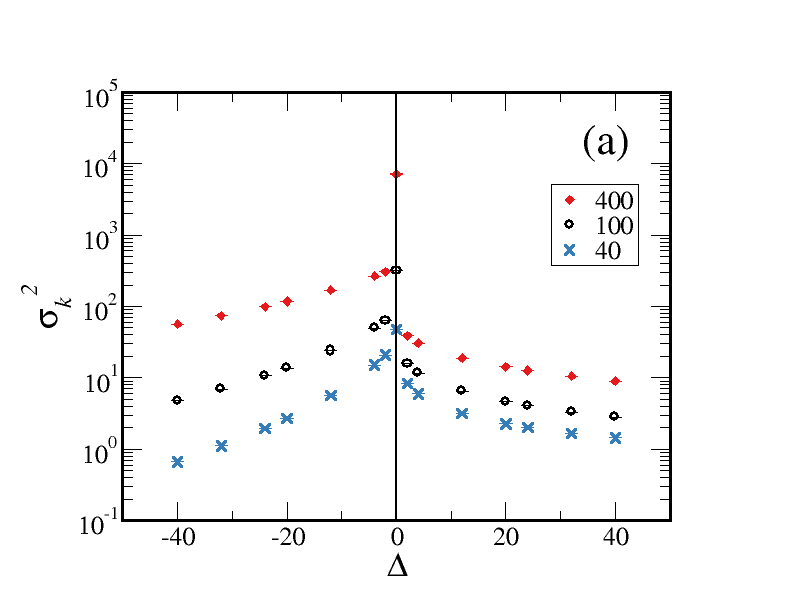} &   \includegraphics[width=80mm]{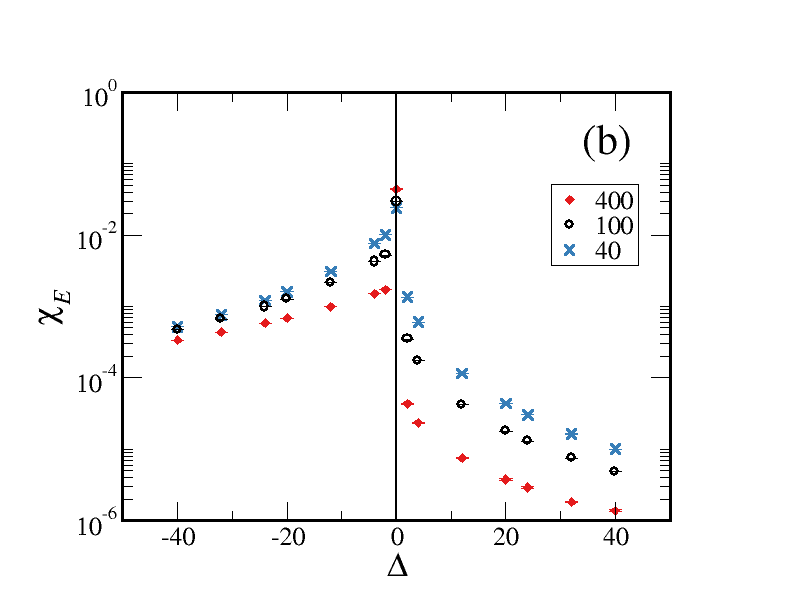} \\
 \includegraphics[width=80mm]{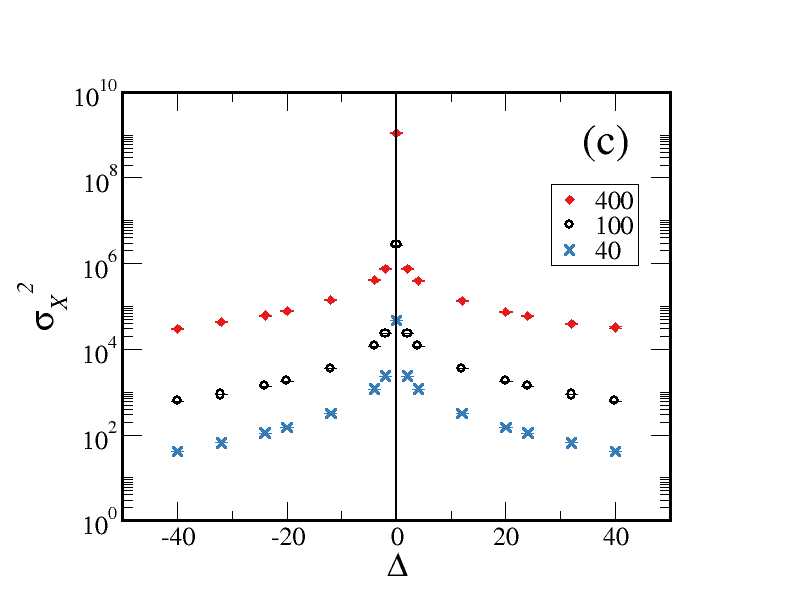} &   \includegraphics[width=80mm]{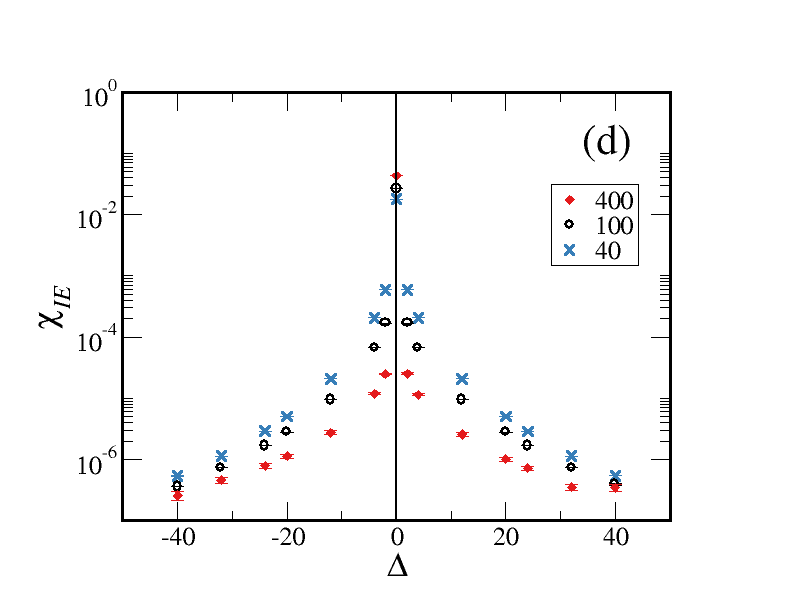} \\
\end{tabular}
    \caption{Simulation data for variances and correlations plotted against $\Delta$. The variances shown are for (a) the introvert
    degree distribution, $\sigma^2_k$, and (c) the total number of cross-links, $\sigma^2_X$. The correlations shown are for
    two links with only the E's being distinct, $\chi_E$, (b) and (d) both the I's and E's being different, $\chi_{IE}$.}
    \label{FlucCorr-vs-Delta}

\end{figure*}

Although the data in Fig. \ref{FlucCorr-vs-Delta} seem peculiar at first
glance, we are able to understand all the unusual properties in two steps.
First, we derived identities which relate all distinct two-point
correlations and the various variances, so that we need to focus our
theoretical considerations on only, say, the $\sigma ^{2}$'s. For the degree
distributions, we exploit a version of the self-consistent mean field
approach (SCMF), first introduced in Ref. \cite{bassler2015extreme}, which
has been improved in several aspects. From its predictions for $\rho \left(
k\right) $, we arrive at a good understanding of the fluctuations $\sigma
_{k}^{2}$. The second step follows the track in Ref. \cite{ZZEB2018}, which
leads us to an excellent approximation for $P^{\ast }\left( X\right) $ and
therefore, predictions for $\sigma _{X}^{2}$, the fluctuations in $X$.

Ending this Section, we will present data for the joint degree distributions
(e.g., $\mu \left( k,p\right) $) and the implied correlations (e.g., $%
C_{IE}\left( k,p\right) $). Unfortunately, finding a viable approximation
for these proves elusive. We provide a hint at the level of complexity
involved in understanding such joint distributions, by computing the
non-trivial $C_{IE}\left( k,p\right) $ for random bipartite graphs exactly.

\subsection{\label{sec:non-crit}Fluctuations and correlations for non-critical 
systems ($\Delta \neq 0$ and fixed $X$ ensembles)}

We begin with Fig. \ref{FlucCorr-vs-Delta}, which shows two sets of
fluctuations and correlations. 

A casual glance of this figure does not hint at any easily tractable
behavior. Indeed the data appear in general to be quite peculiar, displaying
up to \textit{three} regimes, i.e., $\Delta $ being negative, positive, and
zero. To understand these unusual properties and to develop a more coherent
picture, we first note that, though there are apparent differences between
the fluctuation data and the correlations, they are indeed related by the
identities derived above. Since these are exact relations, there is no need
for us to understand the behavior of both. To be specific, we will focus on
only the fluctuations, as we had mean field theories for the associated
distributions. We should emphasize that our data are in complete agreement
with Eqns. (\ref{FCid},\ref{iFCid},\ref{eFCid}), giving us confidence that a
steady state has been established.

Next, since $\rho $ involves $N_{E}$ binary random variables, we expect the
variance $\sigma _{k}^{2}$ to scale with $N_{E}$. Thus, instead of the raw
data, we plot the `normalized' version: $\sigma _{k}^{2}/N_{E}$in Fig. \ref%
{Var-vs-Delta/f}a. 

\begin{figure*}
\begin{tabular}{cc}
 \includegraphics[width=80mm]{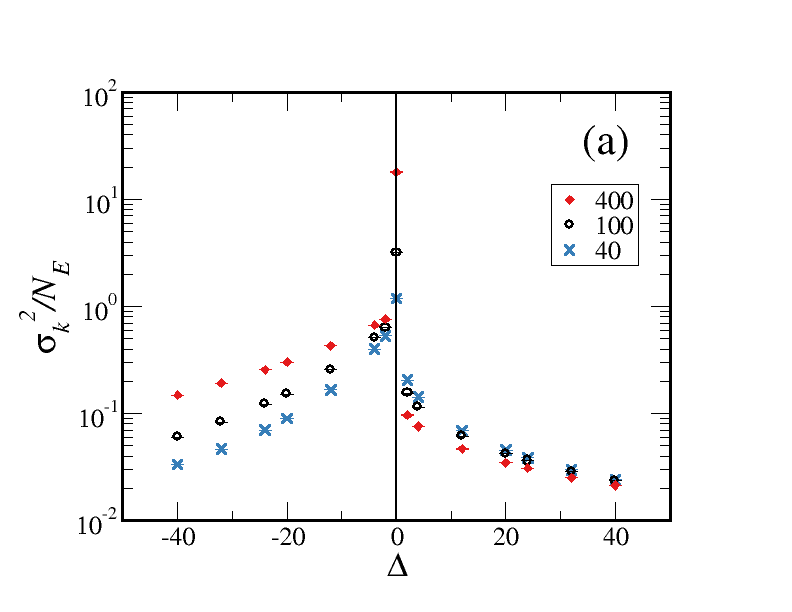} &   \includegraphics[width=80mm]{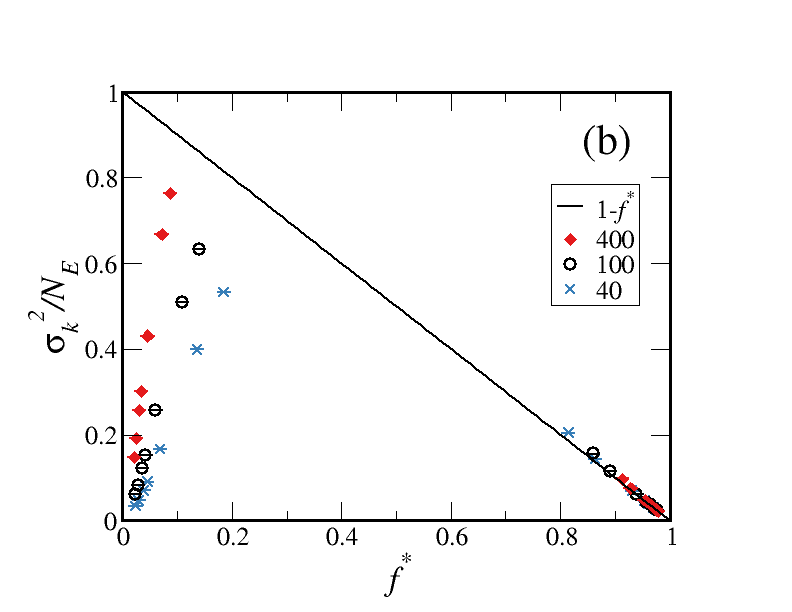} \\
 \includegraphics[width=80mm]{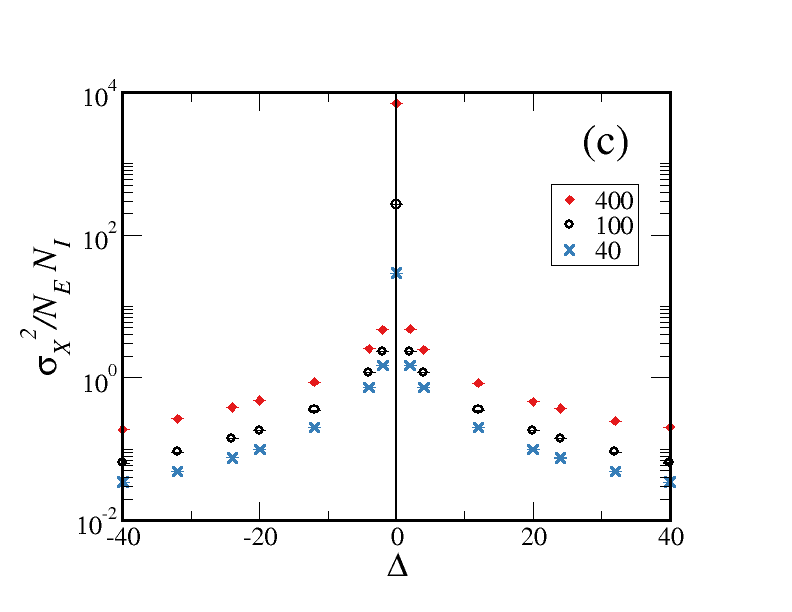} &   \includegraphics[width=80mm]{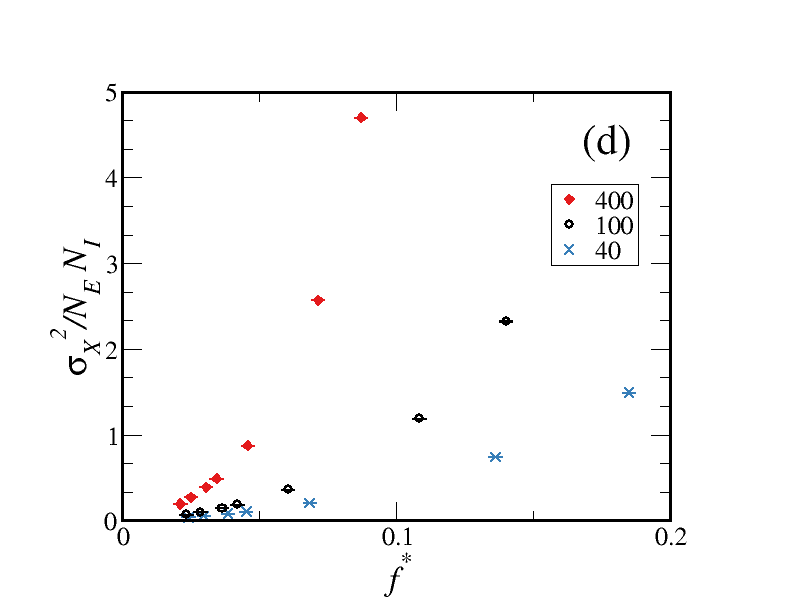} \\
\end{tabular}
    \caption{(a,c) Simulation
data from \ref{FlucCorr-vs-Delta}a,c `normalized' by $N_{E}$ and $\mathcal{N}
$ respectively. (b) Same data set as in (a), except the critical system,
plotted against $f^{\ast }$ instead of $\Delta $. (d) Same data set as in
(b) plotted against $f^{\ast }$; since $\sigma _{X}^{2}$ is symmetric about $%
\Delta =0$, only the small $f^{\ast }$ ($\Delta <0$) region is
displayed.}
\label{Var-vs-Delta/f}

\end{figure*}

\begin{figure}[ht]
     \centering
    \subfloat{\includegraphics[width=0.5\textwidth]{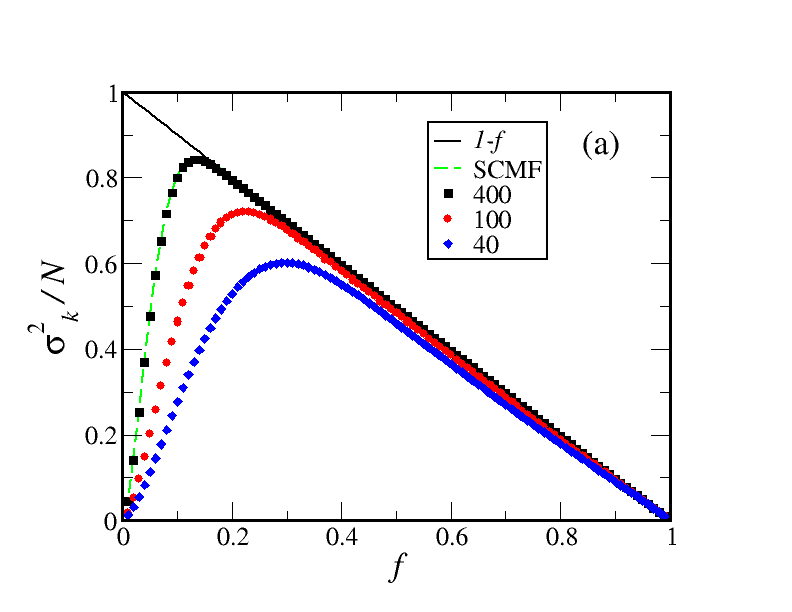}}
    \hspace{-20pt}
    \subfloat{\includegraphics[width=0.5\textwidth]{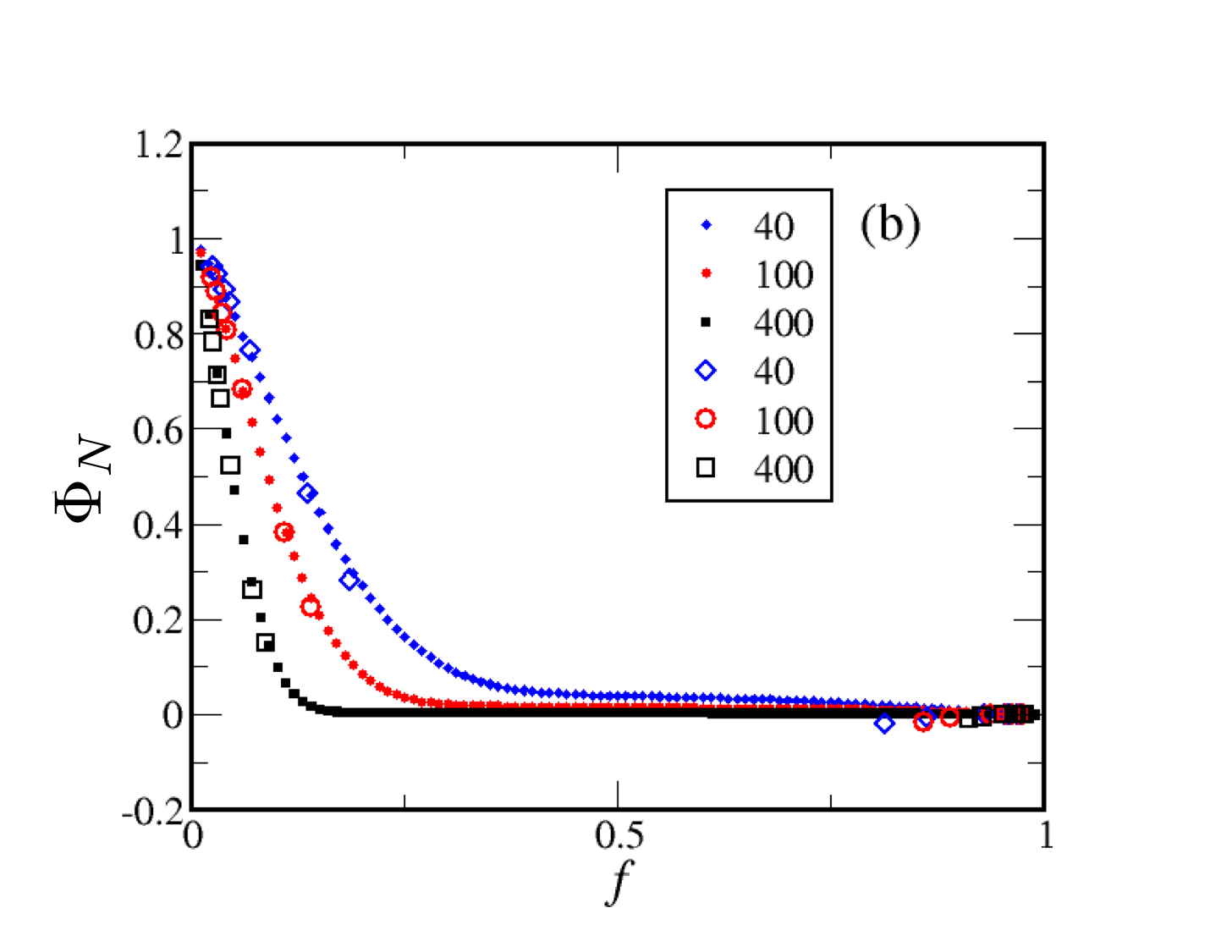}}

    \caption{(a) Simulation data
for  $\widehat{\sigma _{k}^{2}}/N$ (solid symbols) using fixed $X$ (i.e., $f$%
) ensembles for $N=40,100,$ and $400$. As $N$ increases, the data is seen to
converge on $1-f$ (solid line). The dashed line (green online) is the
prediction from our SCMF theory. (b) The difference $\Phi _{N}$ (\ref%
{Phi-def}), highlighting the non-uniform convergence, as $N\rightarrow
\infty $, to a singular step function: $1-\Theta \left( f\right) $. In
addition, $\Phi _{N}$'s from the data in from Fig. \ref{Var-vs-Delta/f}b
(open symbols) are displayed for comparison. See text for discussions.}
\label{Fluc-vs-f}
        
\end{figure}

Similarly, we show $\sigma _{X}^{2}/\mathcal{N}$ in panel
(b). No obvious systematics emerge in this replot. Now, we note that, for
various $N$'s, $f^{\ast }\left( \Delta \right) $ is a rather complicated and
singular function (See Fig. \ref{f-vs-Delta}.). That provides a motivation
for us to plot the variances against $f^{\ast }$: Fig. \ref{Var-vs-Delta/f}%
b,d. Only a minor improvement is seen: tolerable data collapse for $\sigma
_{k}^{2}/N_{E}$ in the $f^{\ast }>1/2$ ($\Delta >0$) regime. Meanwhile, if
we wish to study larger systems, the accessible region of $f^{\ast }$ (with $%
\Delta \neq 0$) becomes more severely limited. In an effort to explore the
`inaccessible' region, we extended our studies to fixed $X$ ensembles of the
critical ($N_{E}=N_{I}$) system. Of course, for such systems, $\widehat{%
\sigma _{X}^{2}}\equiv 0$\ and we are left with only $\widehat{\sigma
_{k}^{2}}$, the data for which are shown in Fig. \ref{Fluc-vs-f}a.

Note that these points are displayed alongside the data from Fig. \ref%
{Var-vs-Delta/f}b, showing that the new points indeed \textquotedblleft fill
in the gap.\textquotedblright\ Though the two sets of data are mostly the
same, there are small differences, the origins of which remain to be
explored further. We conjecture at least two possible sources: (i) Since $f$
is not fixed in the $\Delta \neq 0$ systems (the data point being plotted at 
$f^{\ast }$ here), those fluctuations can contribute to the systematically
larger $\sigma _{k}^{2}/N_{E}$, especially discernible at the $f\sim 1$
regime. (ii) Finite size effects are unlikely to be the same for the two
sets of systems, leading to differences seen in the figure. Finally, we
should mention that there are clear \textquotedblleft
shoulders,\textquotedblright\ especially visible in the $N=40$ data, which
are manifestations of the underlying mesa-like $P^{\ast }\left( X\right) $.
A detailed understanding of such behavior is beyond the scope of this study
and, though quite feasible, will be published
elsewhere.

Turning from these small differences to the more prominent common features,
the most obvious feature is the \textit{non-uniform convergence} of $%
\widehat{\sigma _{k}^{2}}/N$ to the simple function $\left( 1-f\right) $ as $%
N$ increases. One may try to collapse the rise for small $f$ by a naive
function like $Nf$. However, the quality of such collapse is poor \textit{and%
} we cannot find any theoretical justification for such a form. As will be
shown in the next subsection, there exists a non-trivial scaling function
for the difference:%
\begin{equation}
\Phi _{N}\equiv \left( 1-f\right) -\widehat{\sigma _{k}^{2}}/N
\label{Phi-def}
\end{equation}%
Shown in Fig. \ref{Fluc-vs-f}b, $\Phi _{N}$ displays the same non-uniform
convergence properties, but to the singular step function $\Phi _{\infty
}\left( 0\right) =1;\Phi _{\infty }\left( f>0\right) =0$.

Finally, let us return to the fluctuations in $X$ and mention a curious
phenomenon: Dividing $\sigma _{X}^{2}/\mathcal{N}$ by a further factor of $%
N_{I}$, we find reasonable data collapse onto $f^{2}$! There seems to be no
theoretical basis for this behavior and perhaps its appearance is simply `an
accident.' Instead, as the detailed analysis in the next subsection will
show, there is a sound basis for collapse onto a different function, namely, 
$-1-1/f^{\ast }\Delta $.

\subsection{\label{sec:MFAs}Mean field approaches (MFA) and theoretical understanding}

Though the stationary distribution ($\mathcal{P}^{\ast }$) for our model is
explicitly known, it is quite challenging to obtain exact theoretical
results, especially since the system displays an ETE. Surprising progress
had been made, however, through a series of mean field approximations.
Though not systematic, MFAs are based on sound intuitions and captured much
of the essence of the extraordinary properties in our model. Referring the
reader to Refs. \cite{liu2013extraordinary,bassler2015extreme,ZZEB2018} for
the justifications and derivations, we provide only a brief summary here.
The basis of our MFA is the \textit{finite} Poisson distribution (FPD):%
\begin{equation*}
Q\left( n;x,N\right) =\frac{x^{n}}{e_{N}\left( x\right) n!};~~n=0,1,...,N
\end{equation*}%
where $e_{N}\left( x\right) \equiv \sum_{0}^{N}x^{n}/n!$. Since it is rarely
discussed in the literature, we collected in an Appendix a list of its
properties, the most useful of which are $\bar{n}=x\left( 1-Q\left( N\right)
\right) $ and $\overline{n\left( n-1\right) }=x^{2}\left( 1-Q\left( N\right)
-Q\left( N-1\right) \right) $. The FPD enters when we study the \emph{%
introvert} degree distribution in the steady state 
\begin{equation}
\rho \left( k\right) =Q\left( N_{E}-k;\lambda ,N_{E}\right)  \label{Ansatz}
\end{equation}%
as we balance the rate of losing of a link to that for gaining one. In
considering the gain rate, the number of $E$'s \textit{not} already
connected to our $I$ is $N_{E}-k$, while the probability of an $E$ making a
link to our $I$ is a stochastic variable. The parameter $\lambda $ is meant
to capture (the inverse of) the latter probability and so, was argued
previously \cite{ZZEB2018} (where fixed $f$ ensembles were used) to be $%
N_{I}\left( 1-f\right) $. However, when $\bar{k}$ is computed with this $%
\rho $ (and $\lambda $), we find%
\begin{equation}
\bar{k}=N_{E}-\lambda \upsilon  \label{k-bar}
\end{equation}%
where $\upsilon $ is the following function of $\lambda $ 
\begin{equation}
\upsilon \equiv e_{N_{E}-1}\left( \lambda \right) /e_{N_{E}}\left( \lambda
\right) =1-\rho \left( 0\right)  \label{upsilon-def}
\end{equation}%
and represents the fraction of `unsatisfied' $I$'s (those with one or more
links). In other words, given $\lambda $, the FPD will lead to the above $%
\bar{k}$ and therefore a fraction of cross-links being $\bar{k}
/N_{E}=1-\lambda \upsilon $. But if we set $\lambda $ to $N_{I}\left(
1-f\right) $, then this $1-\lambda \upsilon $ can be $f$ only if $\lambda $
takes on a specific value, $\tilde{\lambda}$, which satisfies 
\begin{equation}
\tilde{\upsilon}=N_{E}/N_{I}  \label{upsilon-tilde}
\end{equation}%
where $\tilde{\upsilon}\equiv \upsilon \left( \tilde{\lambda}\right) $. To
emphasize this issue, if we insist on imposing three control parameters ($%
N_{E}$, $N_{I}$, and $f$ -- in fixed $X$ ensembles), then the FPD in Ref. 
\cite{ZZEB2018} \textit{cannot} be a self-consistent approximation in
general. Remarkably, if we do not restrict the value of $f$, then it will
settle, on average, at a $f^{\ast }$ that corresponds to Eqn. (\ref%
{upsilon-tilde}). We will return to this question when we study $P^{\ast
}\left( X\right) $ below.

Here, let us modify the MFA in Ref. \cite{ZZEB2018} to be a self-consistent
mean field theory (SCMF), i.e., we will choose $\lambda $ to be the solution
to 
\begin{equation}
\lambda \upsilon \left( \lambda \right) =N_{E}\left( 1-f\right)
\label{lambda(f)}
\end{equation}%
so that the Ansatz (\ref{Ansatz}) will agree with $\bar{k}=N_{E}f$ in all
cases\footnote{%
We should caution that even this condition cannot be valid for the $XIE$
model in general. After all, if $X$ is fixed, then $\rho \left( k>X\right)
\equiv 0$. If we now impose $X<N_{E}$, then this condition is automatically
violated by the FPD Ansatz. In other words, we should limit our attention to
systems with $f>1/N_{I}$ only.}. Note that, in the context of the FPD, this
relation provides a 1-1 mapping between $f$ and $\lambda $. In our case, it
is straightforward to verify that $df/d\lambda $ is strictly negative.

\begin{figure}[ht]
     \centering
    \subfloat{\includegraphics[width=0.5\textwidth]{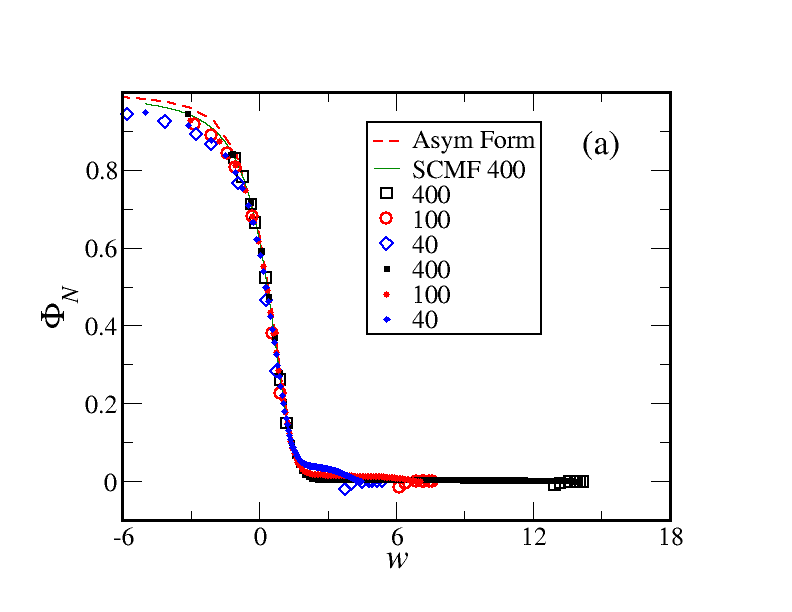}}
    
    \vspace{-20pt}
    \subfloat{\includegraphics[width=0.5\textwidth]{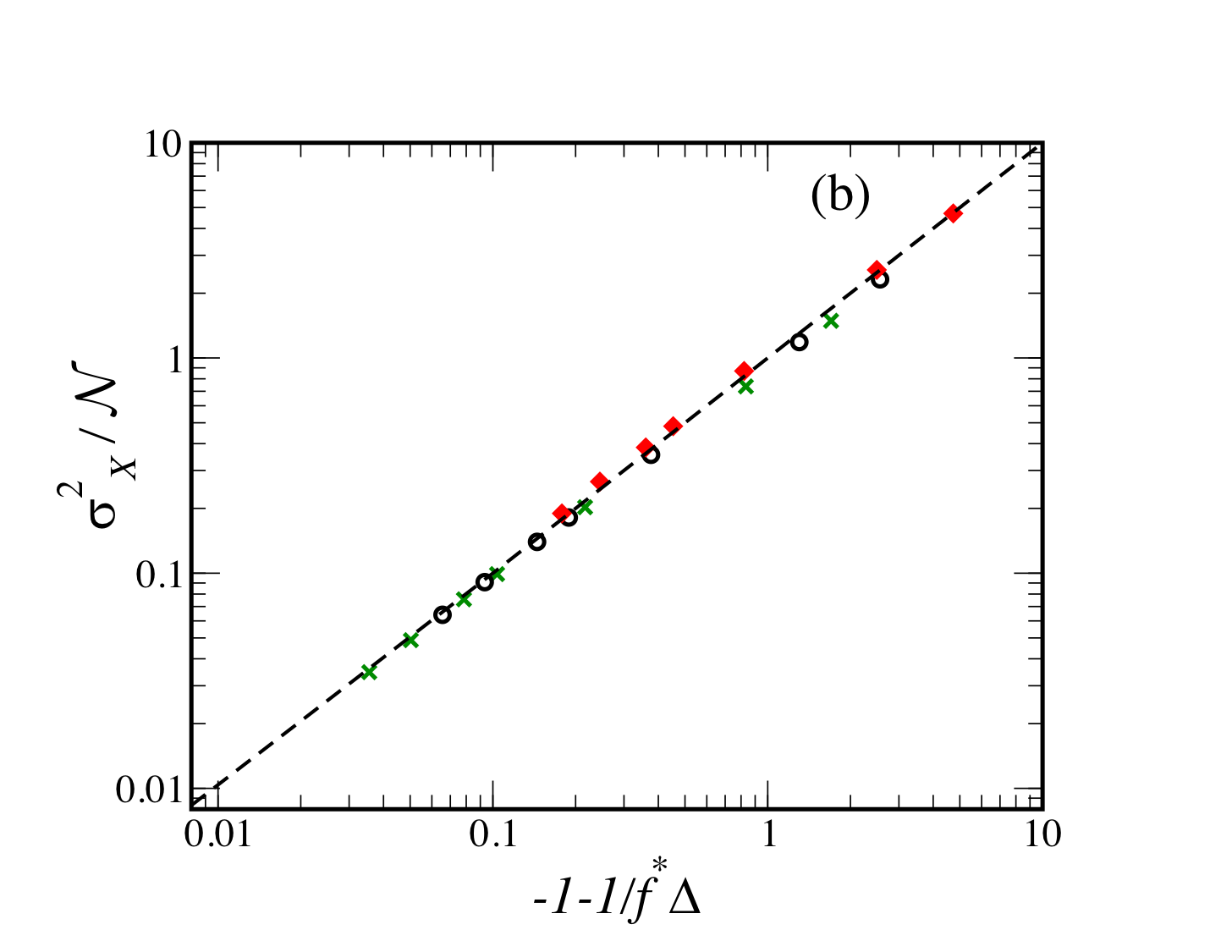}}

    \caption{(a) $\Phi _{N}$ from Fig.\ref{Fluc-vs-f}b (same symbols used)
plotted against the scaling variable $w$, showing good quality data
collapse (especially within $\left( -1,+2\right) $, even for
our small systems). In addition to the data points, we display two curves from SCMF
theory (dash red line for the analytic, asymptotic form (\ref{Phi-final})
and sold black line from numerical evaluation with $N=400$) (b) Log-log plot
of $\sigma _{X}^{2}/\mathcal{N}$ -- the data set in Fig.\ref{Var-vs-Delta/f}%
d -- against the scaling variable $-1-1/f^{\ast }\Delta $. Finite size effects,
especially for the $N=40$ case, restrict the scaling region to a small ragne 
of $w$, in which good quality data collapse is seen over an order of
magnitude (two orders for the $N=400$ data). Dashed line is from numerical
evaluation of the SCMF theory, Eqn. (\ref{scaling-sig2X}). Solid line is the
asymptotic analytic form, Eqn. (\ref{Phi-final}).}
\label{Var-collapse}
        
\end{figure}

Once we grasp the $\lambda $-$f$ connection, we can proceed to compute $%
\widehat{\sigma _{k}^{2}}$ as a function of $f$. In particular, we seek the
non-trivial cross-over behavior for large $N$. As noted above, the
difference $\Phi _{N}$ is more convenient for displaying these properties of 
$\widehat{\sigma _{k}^{2}}$. After some algebra, we find a compact
expression:%
\begin{equation}
\Phi _{N}=\lambda \xi f  \label{Phi=}
\end{equation}%
where%
\begin{equation}
\xi \equiv 1-\upsilon  \label{xi}
\end{equation}%
is the fraction of `satisfied' $I$'s (those no links). The simple form for $%
\Phi _{N}$ is deceptive, however, as both $\lambda $ and $\xi $ are
complicated functions of $f$. The details of the analysis are quite involved
and so, deferred to an Appendix. Here, let us only present the result. As
displayed in Fig.\ref{Fluc-vs-f}, $\Phi _{N}$ drops steeply from $1$ down to
near $0$ at cross-over values which diminishes as $N$ increases.
Nevertheless, as shown in Fig \ref{Var-collapse}a, we find good quality data
collapse (especially within $\left( -1,+2\right) $,
even for systems as small as $40$) 
when this difference is plotted against a scaling variable%
\begin{equation*}
w=\frac{N+1-\lambda }{\sqrt{2\left( N+1\right) }}
\end{equation*}%
which emerges naturally from the analysis of our SCMF theory. As $%
N\rightarrow \infty $, $\Phi _{N}$ is well approximated by the analytic
scaling function\footnote{%
The $\simeq $ sign refers to the right sides as the leading term to an
asymptotic expansion for large $N$. Typically, corrections at $O\left(
N^{-1/2}\right) $ can be expected, as written explicitly\ in many equations
in the Appendix. By contrast, we use $\cong $ to denote a good
approximation, typically unrelated to asymptotic expansions.}%
\begin{equation}
\Phi _{N}\simeq 2\mathcal{E}\left( \mathcal{E}+w\right)  \label{Phi-final}
\end{equation}%
where%
\begin{equation*}
\mathcal{E}\left( w\right) \equiv 1\left/ \sqrt{\pi }e^{w^{2}}\left[ 1+\func{%
erf}\left( w\right) \right] \right.
\end{equation*}%
Both the numerical evaluation of $\Phi _{400}$ and this analytic asymptotic
form are shown in the figure for comparison.

To end this subsection, we turn to the fluctuations in $X$ in non-critical
systems, $\sigma _{X}^{2}/\mathcal{N}$. As shown in Ref. \cite{ZZEB2018}, by
considering the balance of gain and loss of a link in a single attempt, the
equation for the steady state $P^{\ast }\left( X\right) $ is%
\begin{equation*}
N_{E}\left( 1-\zeta _{E}\left( 0\right) \right) P^{\ast }\left( X-1\right)
=N_{I}\left( 1-\rho _{I}\left( 0\right) \right) P^{\ast }\left( X\right)
\end{equation*}%
Again, thanks to $I$-$E$ symmetry, we can focusing on the regime of small $X$
(or $N_{E}/N_{I}<1$), say. Then, we can approximate this equation by%
\begin{equation}
\left( N_{E}/N_{I}\right) P^{\ast }\left( X-1\right) \cong \upsilon P^{\ast
}\left( X\right)  \label{balance}
\end{equation}%
It is clear that, since $\upsilon $ is a monotonic function of $f$ or $X$,
the above recursion relation for $P^{\ast }$ will lead to a maximum
occurring at $\tilde{f}$ (i.e., $\tilde{X}=\tilde{f}\mathcal{N}$) which
satisfies the equation $\tilde{\upsilon}=N_{E}/N_{I}$. To emphasize this
point, the system will be most likely found at $\tilde{f}$ and so, we will
identify it with $f^{\ast }$, the average value of $f$ when the system
settles in the steady state. Note that the condition (\ref{balance})
corresponds to setting the first derivative of $\ln P^{\ast }$ to zero.
Beyond that, both theory and data support the expectation that, as $%
N\rightarrow \infty $, $P^{\ast }\left( X\right) $ approaches a Gaussian
around $\tilde{X}$. Thus, the curvature of $\ln P^{\ast }$ there should
provide us with a good approximation (denoted by $\cong $) for the
fluctuations in $X$. Specifically, we expect%
\begin{eqnarray*}
-1/\sigma _{X}^{2} &\cong &\ln P^{\ast }\left( \tilde{X}+1\right) -2\ln
P^{\ast }\left( \tilde{X}\right) +\ln P^{\ast }\left( \tilde{X}-1\right) \\
&\cong &\ln P^{\ast }\left( \tilde{X}+1\right) -\ln P^{\ast }\left( \tilde{X}%
\right)
\end{eqnarray*}%
In other words, $-1/\sigma _{X}^{2}$ is related to $\left. \partial _{X}\ln
\upsilon \right\vert _{\tilde{X}}$. After some algebra, we arrive at a
concise expression%
\begin{equation}
\sigma _{X}^{2}/\mathcal{N}\simeq \frac{1}{-\tilde{f}\Delta }-1
\label{scaling-sig2X}
\end{equation}%
with corrections of $O\left( N^{-1/2}\right) $ expected. We emphasize that $%
\Delta <0$ for the regime of interest here, so that the first term is
positive. Though we can compute $\tilde{f}$ in terms of the control
parameters $\left( N,\Delta \right) $, we cannot provide a simple expression
as it depends on implicit relations like $\tilde{\upsilon}=N_{E}/N_{I}$.
Nevertheless, we can expect that as $N$ increases with  $| \Delta
| =O\left( 1\right) $, $\tilde{f}$ decreases significantly (with
upper bound $\sqrt{\ln N^{2}/N}$) and so, $\sigma _{X}^{2}/\mathcal{N}$
should become anomalously large. Such a behavior is hardly surprising, as
this limit corresponds to approaching the transition point of our ETE. More
importantly, when all the simulations results are plotted against the
scaling variable, we find high quality data collapse (Fig. \ref{Var-collapse}%
b), even though our $N$'s are not so large. It is reasonable to conclude
that the SCMF here is very successful in capturing the essence of the
anomalous fluctuations in the $XIE$ model.

\subsection{\label{sec:crit}Fluctuations and correlations for critical systems}

\begin{figure}[ht]
\centering
\includegraphics[width=0.5\textwidth]{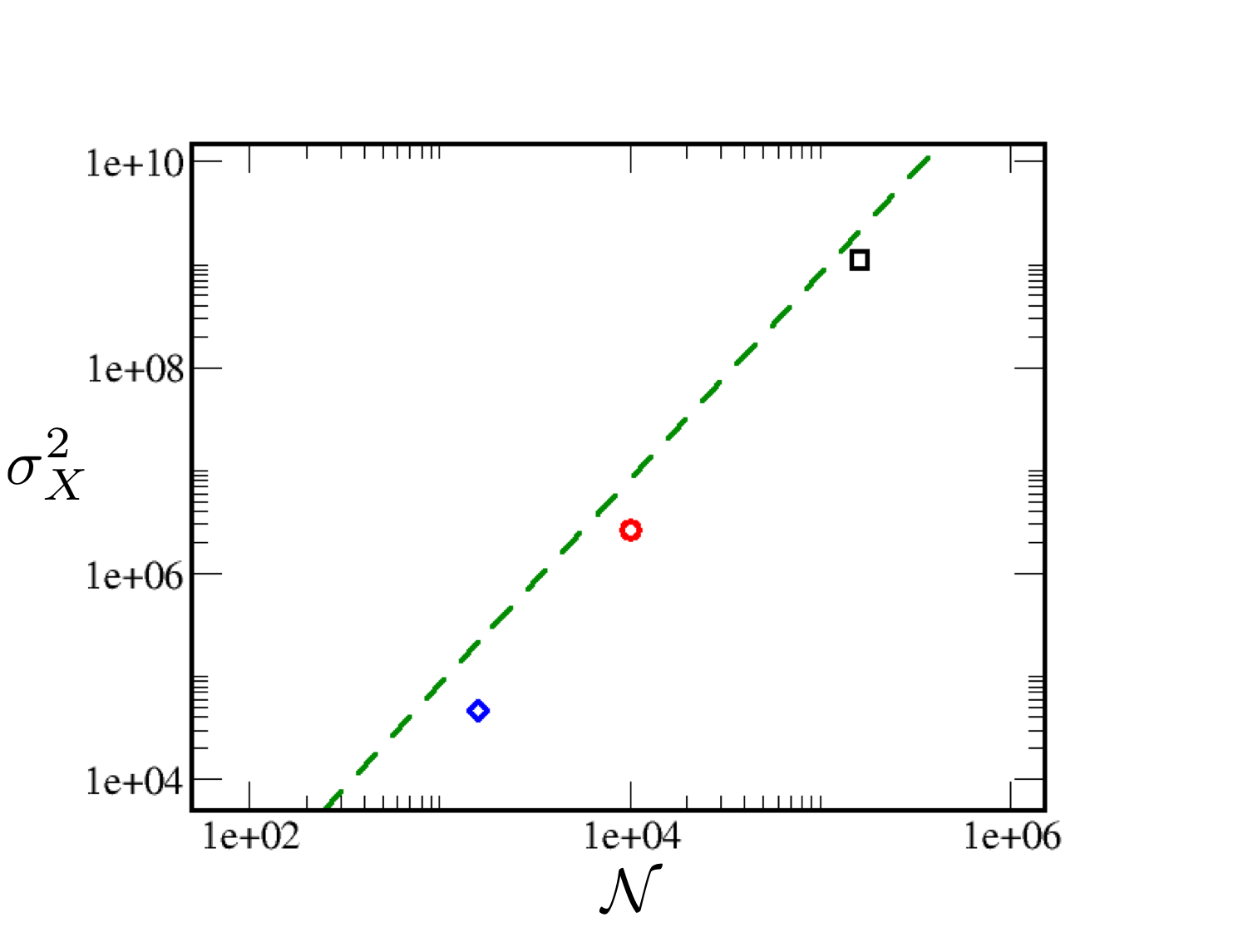}
\caption{Data for $\sigma _{X}^{2}$ (Y axis) in
the three critical systems (symbols) in Fig.\ref{FlucCorr-vs-Delta}c \textit{%
vs. } $\mathcal{N}$ (X axis), along with the (dashed) line $\mathcal{N}^{2}/12$.}
\label{Fluc-crit}
\end{figure}

Finally, we turn to the behavior of critical systems ($\Delta =0$). Here
also, the fluctuation-correlation identities were verified to hold and so,
we again focus our attention only on the former. As Figs \ref%
{FlucCorr-vs-Delta} and \ref{Var-vs-Delta/f} shows, the fluctuations of the
are considerably larger than the ones in the non-critical systems or the
fixed $X$ ensembles. We devote this subsection to a brief summary of how
this extraordinary phenomenon can be understood. We follow the approach in
Ref. \cite{ZZEB2018}, i.e., to regard the critical system as a superposition
of fixed $X$ ensembles, weighted by the critical $P^{\ast }\left(
X;N,0\right) $. Then, for any observable, we have 
\begin{equation*}
\left\langle \mathcal{O}\right\rangle =\sum_{X}\widehat{\left\langle 
\mathcal{O}\right\rangle _{X}}~P^{\ast }\left( X;N,0\right)
\end{equation*}%
For example, this approach predicted $\rho \left( k\right) $ and $\zeta
\left( p\right) $ successfully\cite{ZZEB2018}. Applying it to the
fluctuations of the degree, $\sigma _{k}^{2}$, is straightforward, as we
simply construct the convolution of $\widehat{\sigma _{k}^{2}}$ with $%
P^{\ast }\left( X\right) $. The former is given above, while the latter can
be found in Ref. \cite{ZZEB2018}. Skipping the details, the results for the
three $N$'s are in excellent agreement with data. 
Of course, this approach allows us to understand why
these fluctuations are so much larger than the non-critical ones. While $%
P^{\ast }\left( X;N,\Delta \neq 0\right) $ are Gaussian-like with a
relatively normal spread in $X$, $P^{\ast }\left( X;N,\Delta =0\right) $
resembles a mesa, with sharp drop-offs which approach the boundaries\cite%
{ZZEB2018} as $N\rightarrow \infty $. Indeed, in that limit, there is no
need to compute the location of these drop-offs as $P^{\ast }\left( X\right) 
$ becomes a uniform distribution in $\left[ 0,\mathcal{N}\right] $. Then, $%
\left\langle X^{2}\right\rangle =\mathcal{N}^{2}/3$ and $\sigma _{X}^{2}=%
\mathcal{N}^{2}/12$. In other words, the divergence of Eqn. (\ref%
{scaling-sig2X}) at $\Delta =0$ is bounded by $\mathcal{N}/12$ for large,
but finite, $\mathcal{N}$. Our data for $\mathcal{N}=40^{2},100^{2},400^{2}$
are plotted in Fig. \ref{Fluc-crit}, along with a line showing $\mathcal{N}%
^{2}/12$.

This figure clearly shows that $\sigma _{X}^{2}\propto \mathcal{N}$ is not
satisfied, as it approaches the upper bound for the larger $\mathcal{N}$'s.
To close this subsection, let us point to the analog in the 2D Ising model,
where the fluctuations of the total magnetisation (which corresponds to our $%
X$) of a finite $L\times L$ system diverge anomalously at the critical
point: $\sigma _{M}^{2}\sim L^{4-2\beta /\nu }=\mathcal{N}^{15/8}$.

\subsection{\label{sec:JDD2}Joint degree distributions}

Lastly, we present results of the first explorations into joint
distributions of degrees. For simplicity, we only provide data for $\rho
\left( k,k^{\prime }\right) $ (distribution of degrees of two $I$'s) and $%
\mu \left( k,p\right) $ (degrees of an $I$ and a $E$) in systems with $N=100$%
. To highlight the presence of correlations, we mostly present the
difference between these and the products of the single node degree
distributions, normalized by the latter, i.e., 
\begin{equation}
C_{II}\left( k,k^{\prime }\right) \equiv \frac{\rho \left( k,k^{\prime
}\right) }{\rho \left( k\right) \rho \left( k^{\prime }\right) }%
-1;~~C_{IE}\left( k,p\right) \equiv \frac{\mu \left( k,p\right) }{\rho
\left( k\right) \zeta \left( p\right) }-1  \label{CII+CIE-def}
\end{equation}

\begin{figure}[ht]

\centering
\includegraphics[width=0.45\textwidth]{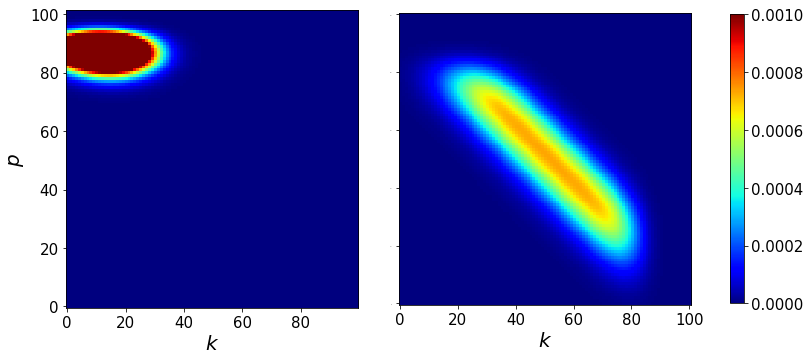}

\includegraphics[width=0.45\textwidth]{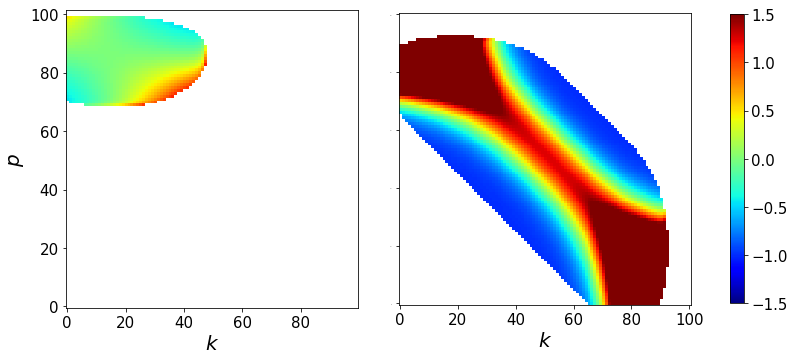}
\caption{Upper panels: The joint degree
distribution $\mu \left( k,p\right) $ for two $N=100$ systems, near
criticality (left, $\Delta =-2$) and at criticality (right, $\Delta =0$).
The elongated region in the latter is a reflection of the mesa-like
distribution, $P^{\ast }\left( X\right) $, at criticality. Lower panels: The
correlations, Eqn. (\ref{CII+CIE-def}), associated with the upper panels.
Note the (highly distorted) quadrapolar form, i.e., correlated near $k=-p$
and anticorrelated close to $k=p$.}
\label{JDD-Delta-XIE-IE}
\end{figure}

\begin{figure}[ht]
\includegraphics[width=0.45\textwidth]{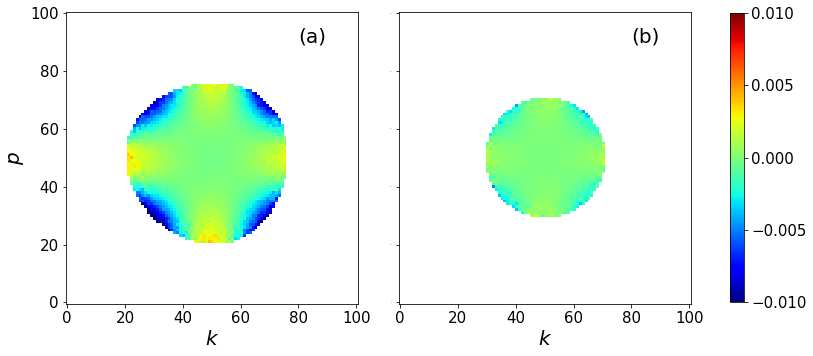}
\caption{Comparison of\ $C_{IE}\left( k,p\right) $\ in two fixed $f=0.5$ ensembles:
$XIE$ (a) and Erd\H{o}s-R\'{e}nyi (b). Note the octaploar patterns in both.}
\label{fXIEvs.fER-IE0.5}
\end{figure}

\begin{figure}[ht]
\includegraphics[width=0.45\textwidth]{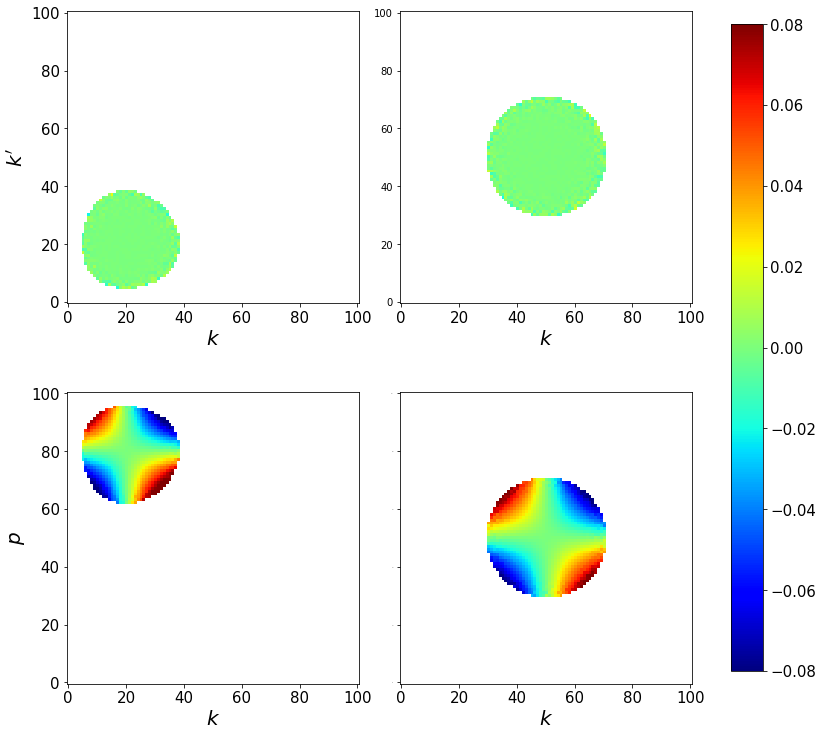}
\caption{Correlations in 
Erd\H{o}s-R\'{e}nyi bipartite ensembles with links being present with
probability $r$. Upper panels: $C_{II}^{ER}\left( k,k^{\prime }\right) $
between two introverts. Lower panels: $C_{IE}^{ER}\left( k,p\right) $,
between an $I$ and an $E$. On the left are ensembles with $r=0.2$ and on the right, $r=0.5$.}
\label{JDD-ER}
\end{figure}

As above, we first consider $XIE$ systems with fixed $\Delta $. To help the
reader visualize the joint distribution itself, we present $\mu \left(
k,p\right) $ for $\Delta =-2$ and $\Delta =0$ in the upper panels of 
Fig. \ref{JDD-Delta-XIE-IE}. As expected from our studies of the
single node distributions $\rho $ and $\zeta $, $\mu $ is mostly narrowly
distributed around $\left( k,p\right) \cong \left( 14,86\right) $ in the
former, but spread over a wide range of values along the line $k+p\cong 50$
in the latter. These qualitative features are shared by the product $\rho
\left( k\right) \zeta \left( p\right) $ of course. The correlations $%
C_{IE}\left( k,p\right) $ are displayed in the lower panels of Fig. \ref%
{JDD-Delta-XIE-IE} and hints at a quadrupolar form: positive along the $k=-p$
diagonal and negative along the $k=p$ line. Though the qualitative aspects
of this phenomenon are easily understood (as they also occur for the 
Erd\H{o}s-R\'{e}nyi random ensemble, Fig. \ref{JDD-ER}d), 
the quantitative features
are more subtle and yet to be examined in detail. Undoubtedly related to the
two-point correlations $\chi $, they remaned to be well understood. Turning
to fixed $X$ ensembles, a more unexpected feature is revealed - octapolar 
$C_{IE}\left( k,p\right) $, as illustrated in Fig. \ref{fXIEvs.fER-IE0.5}a
for a $f=0.5$ system. It is unclear what is the origin of such behavior,
though very small traces of it are visible in the corresponding fixed $X$ 
Erd\H{o}s-R\'{e}nyi ensemble: Fig. \ref{fXIEvs.fER-IE0.5}.

\begin{figure}[ht]
\includegraphics[width=0.45\textwidth]{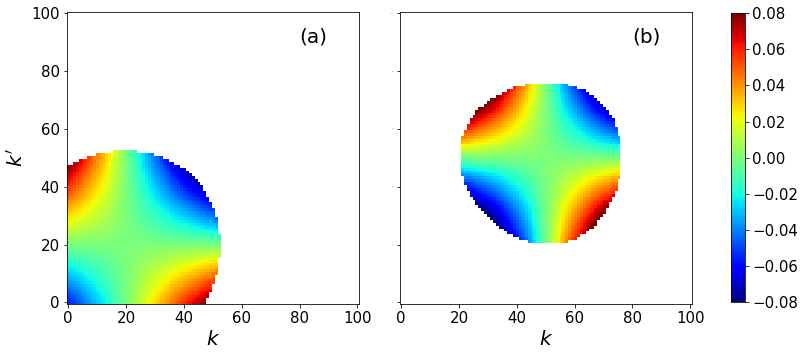}
\caption{Correlations
between two \textit{introverts}, $C_{II}\left( k,k^{\prime }\right) $, in
two fixed $X$ ensembles of $XIE$: (a) $f=0.2$ and (b) $f=0.5$.}
\label{JDD-fXIE-II}
\end{figure}

The next set of figures (\ref{JDD-fXIE-II}) provide
a similar challenge, as they show quadrupolar patterns associated with the joint distribution of two introverts, $C_{II}\left( k,k^{\prime }\right) $,
in fixed $X$ ensembles of the $XIE$ model. Note that, unlike $C_{IE}$, both
variables correspond to degrees (not `holes'). As a result, the $f=0.2$
distribution are centered around $k=k^{\prime }=20$. We conjecture that the
positive/negative correlations along the two diagonals are manifestations of
the fixed $X$ constraint, as increases in $k$ must be compensated by
decreases in $k^{\prime }$. Perhaps there is a deeper connection to the
octapolar pattern in $C_{IE}\left( k,p\right) $. Quantitative analyses are
underway but progress will be challenging, as techniques similar to those
used to build a SCMF for $\rho $ and $\zeta $ fail for joint distributions.

To end this section, we will consider the simpler case of 
Erd\H{o}s-R\'{e}nyi bipartite graphs, i.e., randomly 
distributed cross-links, as the
associated distributions are amenable to analytic tools and can display
non-trivial patterns. First, we study ensembles in which links are present
with a fixed probability,\ $r$. Clearly, we expect 
$C_{II}^{ER}\left(k,k^{\prime }\right) $ to vanish, 
since there is no correlation between two
the entries on different row of $\mathbb{N}$. This property is illustrated
in Fig. \ref{JDD-ER}a,b for $r=0.2$ and $0.5$. By contrast, as shown in Appendix \ref{App_JDER},
we expect $C_{IE}^{ER}\left( k,p\right) $ to be non-trivial, as seen in Fig. \ref{JDD-ER}c,d. We verified that the data are in reasonably good agreement
(i.e., within statistical errors) with the exact quadrupolar form of 
Eqn. (\ref{C-ER}). 
Another class of ER bipartite network is the fixed $f$ ensemble
of random graphs, i.e., all $\mathbb{N}$'s with a fixed fraction ($f$) of
links, each with equal weight. Although we expect the two ensembles to be
equivalent in the $\mathcal{N}\rightarrow \infty $ limit, there appear to be
significant differences, as illustrated in Fig. \ref{fXIEvs.fER-IE0.5}b for 
$C_{IE}^{ER}\left( k,p|X\right) $ in a system with 
$X=\left( 100\right) ^{2}/2$. Not only is the magnitude much smaller than the unconstrained $%
C_{IE}^{ER}\left( k,p\right) $ with $r=0.5$, there is a hint of a octapole
instead of a quadrupole. This puzzling aspect remains to be understood
fully. While further analysis is feasible, it is beyond the scope of this
paper. What we wish to emphasize here is that, even in the case of `simple
ER' bipartite graphs, we find a non-trivial pattern, similar to the one we
found in the constrained $XIE$ model. Clearly, joint distributions offer new
and interesting horizons for future studies.

\section{\label{sec:S+O}Summary and Outlook}

In this study, we examined the correlations and fluctuations in the $XIE$
model. Given that it can be formulated as an 2D Ising model with a
complicated Hamiltonian (which exhibits multi-spin and long-ranged
interactions: Eqn (\ref{Ham})) and that it displayed an extreme Thouless
effect, we should expect serious correlations and fluctuations. Focusing on
two-point correlations, we showed that, in the steady state, the permutation
symmetry of the model implies that there are only three independent
correlations. In $XIE$, these correspond to two links which share one $I$
but connected to two $E$'s ($\chi _{E}$), one $E$ and two $I$'s ($\chi _{I}$%
), or having no common nodes ($\chi _{IE}$). In the Ising language, these
would be two spins on the same row, in the same column, in different rows
and columns. Since there are just three quantities, they can be uniquely
related to three fluctuations: the degrees of $I$ and $E$ and the total
number of cross-links. In the Ising language, these would be the total
magnetisation in a row, a column, and the entire system. We have verified
these fluctuation-correlation identities and chose to focus on the
fluctuations only.

In general, there are two types of fluctuations, associated with
non-critical and critical systems. Drawing an analogy with the 2D Ising
model, we note not only some similarities between the two point correlations
and fluctuations of certain quantities, but also the unusual aspects found
in our system. In all cases, a self-consistent mean-field theory, improved
from previous mean-field approaches, appears to capture the essence of the
Monte Carlo data we obtained. Thus, we conclude that the properties of these
quantities are well understood. An important lesson is that, under the right
circumstances, mean field approaches can be adequate in describing large
fluctuations and strong correlations. Finally, we presented preliminary
studies of joint degree distributions, which offer another perspective into
correlations in the system as well as fresh challenges on how to understand
them.

While the study here provided some insight into the correlations and
fluctuations associated with the extreme Thouless effect in the $XIE$ model,
it also raise natural questions for future research. blah-blah-blah. Beyond
the $XIE$ system, we plan to return to the more generic model of social
networks involving more `realistic' introverts, who prefer few but non-zero
contacts, and extroverts who prefer more but not infinite number of friends.
In general, such systems evolve according to rules that do not obey detailed
balance and so, will settle into non-equilibrium steady states with
non-trivial probability current loops. Thus, we expect studies of these
systems will offer a level of insight into statistical systems not possible
in the equilibrium-like $XIE$ model.

\bigskip

\appendix
\section{Joint distribution for Erd\H{o}s-R\'{e}nyi graphs}
\label{App_JDER}

For Erd\H{o}s-R\'{e}nyi bipartite graphs characterized by $r$ being the
probability for the presence of any link, we have%
\begin{equation*}
\mathcal{P}^{ER}\left( \mathbb{N}\right) =\dprod\limits_{i,\eta }q\left(
n_{i\eta };r\right)
\end{equation*}%
where%
\begin{equation*}
q\left( n_{i\eta };r\right) =r\delta \left( 1-n_{i\eta }\right) +\left(
1-r\right) \delta \left( n_{i\eta }\right)
\end{equation*}%
Thus, 
\begin{eqnarray*}
\rho ^{ER}\left( k\right) &=&\sum_{\mathbb{N}}\delta \left( k-\sum_{\eta
}n_{1\eta }\right) \mathcal{P}^{ER}\left( \mathbb{N}\right) \\
&=&\sum_{\left\{ n_{1\eta }\right\} }\delta \left( k-\sum_{\eta }n_{1\eta
}\right) \dprod\limits_{\eta }q\left( n_{1\eta };r\right) \\
&=&\binom{N_{E}}{k}r^{k}\left( 1-r\right) ^{N_{E}-k}
\end{eqnarray*}%
so that%
\begin{eqnarray*}
\rho ^{ER}\left( k,k^{\prime }\right) &=&\sum_{\mathbb{N}}\delta \left(
k-\sum_{\eta }n_{1\eta }\right) \delta \left( k^{\prime }-\sum_{\eta
}n_{2\eta }\right) \mathcal{P}^{ER}\left( \mathbb{N}\right) \\
&=&\rho ^{ER}\left( k\right) \rho ^{ER}\left( k^{\prime }\right)
\end{eqnarray*}%
is self-evident. Similarly, 
\begin{equation*}
\zeta ^{ER}\left( p\right) =\binom{N_{I}}{p}r^{N_{I}-p}\left( 1-r\right) ^{p}
\end{equation*}%
and $\zeta ^{ER}\left( p,p^{\prime }\right) =\zeta ^{ER}\left( p\right)
\zeta ^{ER}\left( p^{\prime }\right) $. However, for the mixed distribution,
there are only $N_{I}+N_{E}-1$ i.i.d variables, since $n_{11}$ appear in
both $\delta $'s: 
\begin{eqnarray*}
\mu ^{ER}\left( k,p\right) &=&\sum \delta \left( k-\sum_{\eta }n_{1\eta
}\right) \delta \left( p-N_{I}+\sum_{i}n_{i1}\right) \times \\
&&\times q\left( n_{11};r\right) \dprod\limits_{\eta \neq 1}q\left( n_{1\eta
};r\right) \dprod\limits_{i\neq 1}q\left( n_{i1};r\right)
\end{eqnarray*}%
Summing over all but $n_{11}$ first, we have 
\begin{eqnarray*}
\mu ^{ER}\left( k,p\right)  &=&\sum_{n_{11}}q\left( n_{11};r\right) \binom{%
N_{E}-1}{k-n_{11}}r^{k-n_{11}}\left( 1-r\right) ^{N_{E}-1-k+n_{11}}\times  \\
&&\times \binom{N_{I}-1}{p-\left( 1-n_{11}\right) }r^{N_{I}-1-p+1-n_{11}}%
\left( 1-r\right) ^{p-1+n_{11}}
\end{eqnarray*}%
After some algebra, the final result for $\mu ^{ER}\left( k,p\right) /\rho
^{ER}\left( k\right) \zeta ^{ER}\left( p\right) $ can be written as

\begin{eqnarray*}
&&\left[ k\left( N_{I}-p\right) \left( 1-r\right) +\left( N_{E}-k\right) pr%
\right] \frac{1}{\mathcal{N}r\left( 1-r\right) } \\
&=&\left[ \frac{k}{N_{E}}\left( 1-r\right) +\frac{p}{N_{I}}r-\frac{k}{N_{E}}%
\frac{p}{N_{I}}\right] \frac{1}{r\left( 1-r\right) }
\end{eqnarray*}%
If we insert $\bar{k}=rN_{E}$ and $\bar{p}=\left( 1-r\right) N_{I}$, then
the difference from unity can be cast in the simple form%
\begin{equation*}
\frac{\mu ^{ER}}{\rho ^{ER}\zeta ^{ER}}-1=\left( \frac{k}{\bar{k}}-1\right)
\left( 1-\frac{p}{\bar{p}}\right)
\end{equation*}

\section{Finite Poisson distribution (FPD)}
\label{App_FPD}

In this appendix, we provide some properites of this distribution, which
seems to be rarely used, if at all, in the physics literature. Consisting of
a finite number of terms in the standard Poisson distribution, it is the
complement of the truncated Poisson distribution, which is often used in
statistics (e.g., the `zero-truncated Poisson distribution' \cite{Cohen1960}%
). The pdf is defined by%
\[
Q\left( n;x,N\right) \equiv \frac{x^{n}}{e_{N}\left( x\right) n!};~~n\in %
\left[ 0,N\right] 
\]%
where%
\begin{equation}
e_{N}\left( x\right) \equiv \sum_{n=0}^{N}\frac{x^{n}}{n!}  \label{eN-def}
\end{equation}%
is a truncated exponential series ($e_{\infty }\left( x\right) =e^{x}$).
This notation is the same as in Abramowitz and Stegen \cite{A&S1964} (e.g.
6.5.13) and clearly just $e^{x}$ times the cumulative Poisson distribution
function. Thus, it is $e^{x}\Gamma \left( N+1,x\right) /N!$ where $\Gamma $
is the incomplete Gamma function. We list some of its important properties
here.

\begin{enumerate}
\item If $x<N$, $Q$ peaks at $\hat{n}\simeq x$. This is clear, since the
ratio of successive values are $Q\left( n\right) /Q\left( n-1\right) =x/n$.

\item If $x\geq N$, $Q$ is monotonically increasing in $n$. (and peaks at $%
\tilde{n}=N$).

\item Of course, $e_{N}\left( x\right) $ is monotonically increasing in $N$.
Given $x$, $e_{N}\left( x\right) $ has an inflection point at $N\simeq x$
(and saturates to $e^{x}$). It can be regarded as a `partition function' in
that $\ln e_{N}$ plays a major role in computing averages. Note that $%
\partial _{x}^{\ell }e_{N}\left( x\right) =e_{N-\ell }\left( x\right) $.

\item The mean is%
\[
\bar{n}=x\frac{e_{N-1}\left( x\right) }{e_{N}\left( x\right) }=x\partial
_{x}\ln e_{N} 
\]%
Two useful quantities are (i) the last entry of the FPD%
\[
\xi \left( x,N\right) \equiv \frac{x^{N}}{e_{N}\left( x\right) N!} 
\]%
and (ii) its complement%
\[
\upsilon \equiv 1-\xi 
\]%
which is related to $\bar{n}$ via%
\[
\bar{n}=x\upsilon 
\]%
If $x$ is held fixed and $N\rightarrow \infty $, then $e_{N}\rightarrow
e^{x} $ and $\xi \rightarrow 0$ so that $\bar{n}\rightarrow x$. On the other
hand, if keep $x>N$ and we study large $N$, then we must be mindful of $n\in %
\left[ 0,N\right] $ and $\xi $ cannot be small. In this case, it is best to
define a fraction $\bar{n}/N$, or its complement%
\begin{eqnarray}
f &\equiv &1-\bar{n}/N  \label{f-def} \\
&=&1-\left( x/N\right) \partial _{x}\ln e_{N}\left( x\right)  \label{f-eN}
\end{eqnarray}%
for analysis of the large $N$ properties. If $x\gg N$, the simplest result
is obtained by keeping the last few terms in each $\ln e_{N}\simeq N\ln
x+N/x+...$, so that%
\[
\bar{n}/N=1-\frac{1}{x}+... 
\]%
At this lowest order in $N/x$, this result is completely consistent with $%
xf\rightarrow 1$ as $f\rightarrow 0$. The challenge is the cross over around 
$x\sim N$.

\item The second moment is best found from%
\[
\overline{n\left( n-1\right) }=x^{2}\frac{e_{N-2}}{e_{N}}=x^{2}\upsilon
-xN\xi 
\]%
where the last term is actually the next to the last entry of the FPD. From
here, the variance can be obtained. But, the more elegant expression is%
\[
\sigma ^{2}-\bar{n}=x^{2}\partial _{x}^{2}\ln e_{N} 
\]%
Though the explicit expression for $\sigma ^{2}$ is somewhat cumbersome,
there is a simple relationship%
\[
\sigma ^{2}/N=\left( 1-f\right) -x\xi f 
\]%
For fixed $x/N<1$, we have $\xi \rightarrow 0$ exponentially with $%
N\rightarrow \infty $ , so that $\sigma ^{2}/N$ converges to $1-f$. But,
this convergence is not uniform, while the non-trivial, $N$-dependent,
crossover behavior is implicit in $x\xi f$. See Appendix \ref{App_Scaling} for
details.

\item In general, higher moments -- $\overline{n\left( n-1\right) ...\left(
n-\ell +1\right) }=$\ $x^{\ell }\left( \partial _{x}^{\ell }\ln e_{N}\right)
/e_{N}$ -- all vanish for $\ell \geq N$. Thus, averages of any function of $%
n $ can be expressed in terms of the lowest $N$ moments. since the FPD has
only terms up to $N$. Though this may appear strange, we know that, for 
\textit{any} function ($h$) of a variable ($z$) which can take on only
integer values $0,1,...,N$, it can be expressed as a polynomial of degree $N$%
. Denoted by $g_{N}\left( z\right) $, this polynomial depends only on $%
h_{\ell }$, the values of $h$ at integer $\ell \in \left[ 0,N\right] $. In
particular, given a functional form, $h\left( z\right) $, define $%
g_{0}\left( z\right) =h_{0}$ and then, recursively%
\[
g_{\ell }\left( z\right) =g_{\ell -1}\left( z\right) +\frac{h_{\ell
}-g_{\ell -1}\left( \ell \right) }{\ell !}z\left( z-1\right) ...\left(
z-\ell +1\right) 
\]%
for $\ell =1,2,...,N$. For example, $g_{2}\left( z\right) =h_{0}+\left[
h_{1}-h_{0}\right] z+\left[ h_{2}-2h_{1}+h_{0}\right] z\left( z-1\right) /2$%
, and it is easy to check that $g_{2}\left( z\right) $ assumes the values $%
h_{0,1,2}$ for $z=0,1,2$.
\end{enumerate}

\subsection{Large $N$ behavior of $e_{N}\left( x\right) $}
\label{App_largN}
There are three regimes, two are trivial: (i) \textit{fixed} $x$, for which $%
e_{\infty }\left( x\right) =e^{x}$ \ and (ii) $x\gg N$, for which $%
e_{N}\left( x\right) =x^{N}/N!\left[ 1+O\left( N/x\right) \right] $. Our
main interst is the intermediate crossover case of $x\simeq N$, on which
this Appendix is focused.

Using%
\[
\frac{1}{n!}=\frac{1}{2\pi i}\int_{\mathcal{C}}t^{-n}e^{t}\frac{dt}{t} 
\]%
where $\mathcal{C}$ can be any circle around the origin (since $n$ is an
integer), we can rewrite the sum in Eqn. (\ref{eN-def}) as%
\[
e_{N}\left( x\right) =\frac{1}{2\pi i}\int_{\mathcal{C}}\frac{1-\left(
x/t\right) ^{N+1}}{t-x}e^{t}dt 
\]%
Restricting ourselves to $x>0$, we can choose the radius of $\mathcal{C}$\
to be smaller than $x$, so that $\int_{\mathcal{C}}e^{t}/\left( t-x\right)
dt=0$ and we are left with%
\begin{equation}
e_{N}\left( x\right) =\frac{x^{N+1}}{2\pi i}\int_{\mathcal{C}}\frac{t^{-N-1}%
}{x-t}e^{t}dt  \label{int1}
\end{equation}%
This is \textit{exact}, while the large $N$ asymptotics can be extracted by
using the steepest descent method. The saddle point we need lies on the real
axis, at%
\begin{equation}
t_{0}=N+1  \label{t0}
\end{equation}%
and we should deform $\mathcal{C}$\ into the line $t_{0}+iy;y\in \left(
-\infty ,\infty \right) $. So, if $x>t_{0}$, we can ignore the pole.
Otherwise, we must pick up the pole contribution there, which is
\[
e^{x}\Theta \left( N+1-x\right) 
\]%
In general, the discontinuity associated with $\Theta $ must cancel that in
the line integral. Obviously, in regime (i), we can expect this to be the
dominant contribution as $N\rightarrow \infty $.

Turning our attention to the integral over $y$, we make the usual expansion
and, keeping only the lowest non-trivial term in the exponent, find%
\[
\begin{aligned} \frac{t^{-N-1}}{x-t}e^{t}={} \frac{1}{x-t_{0}\left(
1+iy\right) }\exp [ t_{0}-\left( N+1\right) \ln t_{0} \\ -\left(
N+1\right) \frac{y^{2}}{2}-\left( N+1\right) \frac{y^{3}}{3} ... ]
\end{aligned}
\]%
Changing the variable of integration to \\ $\eta =y/\sqrt{2/\left( N+1\right) }$
and defining 
\[
w\equiv \frac{N+1-x}{\sqrt{2\left( N+1\right) }}
\]%
we arrive at%
\begin{equation}
e_{N}\left( x\right) \simeq e^{x}\Theta \left( N+1-x\right) -\frac{1}{2\pi }%
\left( \frac{xe}{N+1}\right) ^{N+1}\int_{-\infty }^{\infty }\frac{e^{-\eta
^{2}}d\eta }{w+i\eta }  \label{eN2}
\end{equation}%
Note that the integral is singular at $w=0$. To continue, we exploit%
\[
\frac{1}{w+i\eta }=\int_{0}^{\infty }du\left\{ 
\begin{array}{cc}
e^{-u\left( w+i\eta \right) } & \text{for }w>0 \\ 
-e^{-u\left( \left\vert w\right\vert -i\eta \right) } & \text{for }w<0%
\end{array}%
\right. 
\]%
perform the $\eta $ integration, and define%
\begin{equation}
\mathcal{J}\left( \left\vert w\right\vert \right) \equiv \frac{1}{2\sqrt{\pi 
}}\int_{0}^{\infty }due^{-u^{2}/4-u\left\vert w\right\vert }=\frac{1}{2}%
e^{w^{2}}\limfunc{erfc}\left( \left\vert w\right\vert \right)   \label{J-def}
\end{equation}%
where $\limfunc{erfc}$ is the complementary error function. The result is 
\begin{equation}
e_{N}\left( x\right) \simeq e^{x}\Theta \left( w\right) -\left( \frac{xe}{N+1%
}\right) ^{N+1}\limfunc{sgn}\left( w\right) \mathcal{J}\left( \left\vert
w\right\vert \right)   \label{final}
\end{equation}%
As $w\rightarrow 0_{\pm }$, the second term suffers a discontinuity of $%
e^{N+1}$ and cancels that in the first term. Though this form will turn out
to be more convenient (especially for considering the $w<0$ regime), we can
write (\ref{final}) in a way that the discontinuity is manifestly zero
(since $1-2\mathcal{J}\left( 0\right) =0$). First, to keep quantities to $%
1+O\left( N^{-1/2}\right) $, we will need%
\begin{eqnarray}
\left( \frac{x}{N+1}\right) ^{N+1} =\left( 1-\frac{w\sqrt{2\left(
N+1\right) }}{N+1}\right) ^{N+1} &\\
=e^{-w\sqrt{2\left( N+1\right) }-w^{2}}\left( 1+O\left( N^{-1/2}\right)
\right)   \label{x^N+1}&
\end{eqnarray}%

so that the second term in (\ref{final}) becomes $e^{x}\limfunc{sgn}\left(
w\right) \limfunc{erfc}\left( \left\vert w\right\vert \right) /2$. The
result is 
\begin{equation}
e_{N}\left( x\right) \simeq e^{x}\left[ 1+\func{erf}\left( w\right) \right]
/2  \label{final-pretty}
\end{equation}%
where we have used $\Theta \left( w\right) -\limfunc{sgn}\left( w\right) 
\limfunc{erfc}\left( \left\vert w\right\vert \right) /2=\left[ 1+\func{erf}%
\left( w\right) \right] /2$ for $w>0$ and, for $w<0$, $\limfunc{erfc}\left(
-w\right) =1-\func{erf}\left( -w\right) =1+\func{erf}\left( w\right) $. It
can be found in Ref. \cite{A&S1964}, 26.4.11.

The most crucial result of this analysis is the identification of the
scaling variable, $w$. Not surprisingly, $w=0$ is associated with the
crossing over of the FPD from being an ordinary PD to a Laplacian like
distribution peaked at $n=N$. In addition, at $w=0$, we find $f=\xi +O\left(
1/N\right) =\sqrt{2/\left( \pi N\right) }+O\left( 1/N\right) $, a value
which fits well into where we observe the cross overs in the data.

Before ending this section, let us point out an easier route to an
approximation which is also quite good for large $N$. This approach relies
on approximating the (standard) Poisson distribution by a Gaussian for large 
$x$, i.e., 
$e^{-x}x^{n}/n!\simeq \left[ 2\pi x\right] ^{-1/2}\exp \left\{ -\frac{\left( n-x\right) ^{2}}{2x}\right\} $. 
The cumulative distribution of
the latter (by regarding $n$ as a continuous variable) is just 
$F\left(n;x\right) \equiv \left[ 1+\func{erf}\left\{ \left( n-x\right) /\sqrt{2x}\right\} \right] /2$, 
leading to the approximate expression 
$e_{N}\left( x\right) \simeq e^{x}F\left( N;x\right) $. It is clear that this form
differs from Eqn. (\ref{final-pretty}) by $O\left( 1/N\right) $ for large $N$. 

\subsection{Scaling form for the crossover function $\Phi _{N}$}
\label{App_Scaling}
Here, we present details for finding the scaling form for 
\[
\Phi _{N}=x\xi f
\]%
First, using (\ref{f-def}), we can write $\Phi $ in terms of $x\xi /\sqrt{N}$%
:%
\[
\Phi _{N}=x\xi \left( 1-\frac{x}{N}+\frac{x\xi }{N}\right) \simeq \sqrt{2}w%
\frac{x\xi }{\sqrt{N}}+\left( \frac{x\xi }{\sqrt{N}}\right) ^{2}
\]%
Next, (\ref{final},\ref{x^N+1}), we have, for $w<0$,%
\begin{eqnarray*}
&\frac{\sqrt{N}}{x\xi }=\frac{e_{N}\left( x\right) \sqrt{N}}{x^{N+1}/N!} &\\%
&\simeq \left. \sqrt{2\pi }\left( \frac{xe}{N+1}\right) ^{N+1}\mathcal{J}%
\left( -w\right) \right/ \left( \frac{xe}{N+1}\right) ^{N+1}& \\
&=\sqrt{\pi /2}e^{w^{2}}\limfunc{erfc}\left( -w\right) \left( 1+O\left(
N^{-1/2}\right) \right) &
\end{eqnarray*}%


and for $w>0$, 
\begin{eqnarray*}
\frac{\sqrt{N}}{x\xi } &\simeq &\left. \sqrt{2\pi }e^{x}\right/ \left( \frac{%
xe}{N+1}\right) ^{N+1}-\sqrt{\pi /2}e^{w^{2}}\limfunc{erfc}\left( w\right) 
\\
&\simeq &\sqrt{\pi /2}e^{w^{2}}\left[ 2-\limfunc{erfc}\left( w\right) \right]
\end{eqnarray*}%
Since $\limfunc{erfc}\left( -w\right) =1-\func{erf}\left( -w\right) =1+\func{%
erf}\left( w\right) $, both of these can be combined into a single compact
form. The result is, explicitly,%
\[
\Phi _{N}=2\mathcal{E}\left( \mathcal{E}+w\right) \left( 1+O\left(
N^{-1/2}\right) \right) 
\]%
where $\mathcal{E}\left( w\right) \equiv 1\left/ \sqrt{\pi }e^{w^{2}}\left[
1+\func{erf}\left( w\right) \right] \right. $.


\begin{thebibliography}{10}

\bibitem{kafri_why_2000}
Yariv Kafri, David Mukamel, and Luca Peliti.
\newblock Why is the {DNA} {Denaturation} {Transition} {First} {Order}?
\newblock {\em Physical Review Letters}, 85(23):4988--4991, December 2000.

\bibitem{poland_phase_1966}
Douglas Poland and Harold~A. Scheraga.
\newblock Phase {Transitions} in {One} {Dimension} and the {Helix}—{Coil}
  {Transition} in {Polyamino} {Acids}.
\newblock {\em The Journal of Chemical Physics}, 45(5):1456--1463, September
  1966.

\bibitem{fisher_effect_1966}
Michael~E. Fisher.
\newblock Effect of {Excluded} {Volume} on {Phase} {Transitions} in
  {Biopolymers}.
\newblock {\em The Journal of Chemical Physics}, 45(5):1469--1473, September
  1966.

\bibitem{Thouless_long-range_1969}
DJ~Thouless.
\newblock Long-{Range} {Order} in {One}-{Dimensional} {Ising} {Systems}.
\newblock {\em Physical Review}, 187(2):732--733, November 1969.

\bibitem{BarMukamel}
Amir Bar and David Mukamel.
\newblock Mixed-{Order} {Phase} {Transition} in a {One}-{Dimensional} {Model}.
\newblock {\em Physical Review Letters}, 112(1):015701, January 2014.

\bibitem{FFK16}
Agata Fronczak, Piotr Fronczak, and Andrzej Krawiecki.
\newblock Minimal exactly solved model with the extreme thouless effect.
\newblock {\em Physical Review E}, 93(1):012124, 2016.

\bibitem{liu2013modeling}
Wenjia Liu, Shivakumar Jolad, Beate Schmittmann, and RKP Zia.
\newblock Modeling interacting dynamic networks: I. preferred degree networks
  and their characteristics.
\newblock {\em Journal of Statistical Mechanics: Theory and Experiment},
  2013(08):P08001, 2013.

\bibitem{Note1}
We exclude all self-contacts: $a_{ii}\equiv 0$.

\bibitem{zia2007probability}
RKP Zia and B~Schmittmann.
\newblock Probability currents as principal characteristics in the statistical
  mechanics of non-equilibrium steady states.
\newblock {\em Journal of Statistical Mechanics: Theory and Experiment},
  2007(07):P07012, 2007.

\bibitem{liu2014modeling}
Wenjia Liu, B~Schmittmann, and RKP Zia.
\newblock Modeling interacting dynamic networks: Ii. systematic study of the
  statistical properties of cross-links between two networks with preferred
  degrees.
\newblock {\em Journal of Statistical Mechanics: Theory and Experiment},
  2014(5):P05021, 2014.

\bibitem{zia2012extraordinary}
RKP Zia, Wenjia Liu, and B~Schmittmann.
\newblock An extraordinary transition in a minimal adaptive network of
  introverts and extroverts.
\newblock {\em Physics Procedia}, 34:124--127, 2012.

\bibitem{liu2013extraordinary}
Wenjia Liu, Beate Schmittmann, and RKP Zia.
\newblock Extraordinary variability and sharp transitions in a maximally
  frustrated dynamic network.
\newblock {\em EPL (Europhysics Letters)}, 100(6):66007, 2013.

\bibitem{bassler2015extreme}
KE~Bassler, Wenjia Liu, B~Schmittmann, and RKP Zia.
\newblock Extreme thouless effect in a minimal model of dynamic social
  networks.
\newblock {\em Physical Review E}, 91(4):042102, 2015.

\bibitem{bassler2015networks}
Kevin~E Bassler, Deepak Dhar, and RKP Zia.
\newblock Networks with preferred degree: a mini-review and some new results.
\newblock {\em Journal of Statistical Mechanics: Theory and Experiment},
  2015(7):P07013, 2015.

\bibitem{ZZEB2018}
RKP Zia, Weibin Zhang, Mohammadmehdi Ezzatabadipour, and Kevin~E Bassler.
\newblock Exact results for the extreme thouless effect in a model of network
  dynamics.
\newblock {\em EPL (Europhysics Letters)}, 124(6):60008, 2018.

\bibitem{Note2}
There is considerable work on adaptive networks in the literature, especially
  in connection with epidemic spreading\cite {GrossGeneral}. However, nearly
  all other approaches are based on rewiring existing links from, e.g., an
  infected node to a healthy one. By introducing \protect \textit {preferred
  degrees}, we believe our models capture human behavior more realistically, as
  well as the possibility of diverse and inhomogeneous population.

\bibitem{platini2010}
T~Platini and RKP Zia.
\newblock Network evolution induced by the dynamical rules of two populations.
\newblock {\em Journal of Statistical Mechanics: Theory and Experiment},
  2010(10):P10018, 2010.

\bibitem{jolad2012}
Shivakumar Jolad, Wenjia Liu, Beate Schmittmann, and RKP Zia.
\newblock Epidemic spreading on preferred degree adaptive networks.
\newblock {\em PloS one}, 7(11):e48686, 2012.

\bibitem{Note3}
Unless stated otherwise, Latin/Greek indices are reserved for
  introverts/extroverts.

\bibitem{Note4}
We denote quantities associated with the stationary state by a superscript
  ($^{\ast }$).

\bibitem{Note5}
For example, $\protect \qopname \relax o{ln}\left ( n_{1}+n_{2}+n_{3}\right )
  !=\left ( n_{1}n_{2}+n_{2}n_{3}+n_{3}n_{1}\right ) \protect \qopname \relax
  o{ln}2+n_{1}n_{2}n_{3}\protect \qopname \relax o{ln}\left ( 3/4\right ) $.

\bibitem{Note6}
To be consistent with our notation, we should write $\left < ... \right >
  ^{\ast }$ for stationary averages. But, for the sake of simplicity, we drop
  the superscript, since all averages considered in this paper will be in the
  stationary state. By contrast, note that, e.g., while $f^{\ast }\equiv \left
  < f\right > $ is a single number, $f$ denotes a variable.

\bibitem{Note7}
Again, we drop the superscript $^{\ast }$ for $\rho $ for simplicity, despite
  that it is a steady state quantity.

\bibitem{Note8}
Note the bar above denote averages over the degree distributions, not the
  microscopic $\protect \mathcal {P}\left ( \protect \mathbb {N}\right ) $.

\bibitem{SZ95}
B~Schmittmann and RKP Zia.
\newblock Statistical mechanics of driven diffusive systems.
\newblock {\em Phase transitions and critical phenomena}, 17:3--214, 1995.

\bibitem{DZ18}
Ronald Dickman and RKP Zia.
\newblock Driven widom-rowlinson lattice gas.
\newblock {\em Physical Review E}, 97(6):062126, 2018.

\bibitem{E-R}
P.~Erdos and A.~Renyi.
\newblock On random graphs i.
\newblock {\em Publicationes Mathematicae}, 6:290--297, 1959.

\bibitem{Note9}
To be clear, the full notation should be denoted $P^{\ast }\left ( X;N,\Delta
  \right ) $, showing the dependence on the control parameters $\left (
  N,\Delta \right ) $. For simplicity, we suppress the latter except when their
  presence are crucial.

\bibitem{YangLee52}
CN~Yang and TD~Lee.
\newblock Statistical theory of equations of state and phase transitions. ii.
  lattice gas and ising model.
\newblock {\em Physical Review}, 87(6):410--419, 1952.

\bibitem{Note10}
The superscript, $ER$, is remind us that this result only holds for Erd\H
  {o}s-R\'{e}nyi graphs.

\bibitem{Note11}
If we write $\left ( k-\protect \mathaccentV {bar}016{k}\right ) =\protect
  \mathaccentV {bar}016{k}\protect \qopname \relax o{cos}\theta $ and $\left (
  p-\protect \mathaccentV {bar}016{p}\right ) =\protect \mathaccentV
  {bar}016{p}\protect \qopname \relax o{sin}\theta $, then we find the standard
  quadrupole form: $C\propto \protect \qopname \relax o{sin}2\theta $.

\bibitem{Note12}
We should caution that even this condition cannot be valid for the $XIE$ model
  in general. After all, if $X$ is fixed, then $\rho \left ( k>X\right ) \equiv
  0$. If we now impose $X<N_{E}$, then this condition is automatically violated
  by the FPD Ansatz. In other words, we should limit our attention to systems
  with $f>1/N_{I}$ only.

\bibitem{Note13}
The $\simeq $ sign refers to the right sides as the leading term to an
  asymptotic expansion for large $N$. Typically, corrections at $O\left (
  N^{-1/2}\right ) $ can be expected, as written explicitly\ in many equations
  in the Appendix. By contrast, we use $\protect \cong $ to denote a good
  approximation, typically unrelated to asymptotic expansions.

\bibitem{Cohen1960}
A~Clifford Cohen.
\newblock Estimating the parameter in a conditional poisson distribution.
\newblock {\em Biometrics}, 16(2):203--211, 1960.

\bibitem{A&S1964}
Milton Abramowitz and Irene Stegun, editors.
\newblock {\em Handbook of Mathematical Functions with Formulas, Graphs, and
  Mathematical Tables}.
\newblock United States Department of Commerce, National Bureau of Standards,
  1964.

\bibitem{GrossGeneral}
T~Gross and H~Sayama, editors.
\newblock {\em Adaptive Networks: Theory, Models and Applications}.
\newblock NECSI/Springer, 2009.

\end{thebibliography}

\end{document}